\documentclass[paper]{JHEP3} 

\usepackage{epsfig,multicol}
\usepackage{amsmath}
\usepackage{amscd}
\usepackage{amsfonts}
\usepackage{amsthm}

\def\IC{\mathbb{C}}
\def\IN{\mathbb{N}}
\def\IZ{\mathbb{Z}}
\def\IR{\mathbb{R}}
\def\ID{\mathbb{D}}

\def\IH{\mathbb{H}}
\def\IF{\mathbb{F}}

\newcommand{\pp}{{=\!\!\!|}}
\newcommand\fverb{\setbox\pippobox=\hbox\bgroup\verb}
\newcommand\fverbdo{\egroup\medskip\noindent%
            \fbox{\unhbox\pippobox}\ }
\newcommand\fverbit{\egroup\item[\fbox{\unhbox\pippobox}]}
\newbox\pippobox

\title{An $N=2$ worldsheet approach to D-branes in bihermitian geometries: I. Chiral and twisted
chiral fields}

\author{Alexander Sevrin, Wieland Staessens\thanks{Aspirant FWO.}\\
Theoretische Natuurkunde, Vrije Universiteit Brussel and\\The International Solvay Institutes\\
Pleinlaan 2, B-1050 Brussels, Belgium \\
E-mail:  \email{Alexandre.Sevrin@vub.ac.be},
\email{Wieland.Staessens@vub.ac.be}}
\author{Alexander Wijns\\Science Institute, University of Iceland,\\
Dunhaga 3, 107 Reykjav\'\i k, Iceland\\
E-mail: \email{awijns@raunvis.hi.is}}

\abstract{We investigate $N=(2,2)$ supersymmetric nonlinear $\sigma$-models in the presence of a
boundary. We restrict our attention to the case where the bulk geometry is described by chiral and twisted chiral
superfields corresponding to a bihermitian bulk geometry with two commuting complex structures. The D-brane configurations
preserving an $N=2$ worldsheet supersymmetry are identified. Duality transformations interchanging chiral for twisted
chiral fields and vice versa while preserving all supersymmetries are explicitly
constructed. We illustrate our results with various explicit examples such as the WZW-model on the
Hopf surface $S^3\times S^1$.
The duality transformations provide e.g new examples of coisotropic A-branes on K{\"a}hler manifolds
(which are not necessarily hyper-K{\"a}hler). Finally, by dualizing a chiral and a twisted chiral field to a
semi-chiral multiplet, we initiate the study of D-branes in bihermitian geometries where the cokernel of the
commutator of the complex structures is non-empty.}


\begin{document}
\setcounter{equation}{0}

%
%

\section{Introduction} \label{introduction}
Non-linear $ \sigma $-models in two dimensions with an $N=(2,2)$
supersymmetry, \cite{Alvarez-Gaume:1981hm}--\cite{Howe:1985pm}, are
an important tool in the study of type II superstrings in the
absence of R-R fluxes.
The (local) target space geometry of such models is characterized by a metric,
a closed 3-form and two complex structures.
The complex structures are covariantly constant and the metric is hermitian with respect to both
complex structures. These conditions can be solved in terms of a single real potential,
\cite{Lindstrom:2005zr} (building on results in \cite{Buscher:1987uw}--\cite{Bogaerts:1999jc}),
which has a
natural interpretation in $d=2$, $N=(2,2)$ superspace where it is the Lagrange density.
The Lagrange density is a
function of three types of scalar superfields (satisfying certain constraints linear in the
superspace derivatives) \cite{Lindstrom:2005zr}, \cite{MS}: chiral, twisted chiral
and semi-chiral superfields.

Whenever one wants to deal with (open) strings propagating in backgrounds
which include D-branes one necessarily needs to confront $N=(2,2)$
non-linear $ \sigma $-models with boundaries. Having a boundary breaks
the $N=(2,2)$ supersymmetry down to an $N=2$ supersymmetry. While a lot
of attention has been devoted to these models
\cite{Ooguri:1996ck}--\cite{Lindstrom:2002jb}, their full description in
$N=2$ superspace remained till recently an unstudied problem. An initial
investigation in \cite{Koerber:2003ef} showed that this was
straightforward as long as one only deals with chiral fields or put
differently as long as one considers B-branes on K{\"a}hler manifolds. In
\cite{Sevrin:2007yn} this was extended to A-branes on K{\"a}hler manifolds.
The field content of these models consists exclusively of twisted chiral
fields. The treatment of twisted chiral fields in $N=2$ boundary
superspace turned out to be rather subtle and an elegant and rich
structure emerged. Duality transformations turning A- into B-branes and
vice-versa were developed as well.

K{\"a}hler manifolds are only a particular example of the geometries which
allow for an $N=(2,2)$ bulk supersymmetry. In general such a geometry is
called bihermitian. In the present paper we extend the analysis of
\cite{Sevrin:2007yn} to bihermitian geometries restricting ourselves to
the simplest non-trivial case in which the two complex structures
associated with the bihermitian geometry mutually commute. In $N=(2,2)$
superspace this corresponds to the case in which the bulk geometry is
parameterized by chiral and twisted chiral superfields simultaneously.
While still relatively simple, these models already encompass the K{\"a}hler
case as they allow for non-trivial NS-NS backgrounds.

Finally, let us remark that the study of D-branes in the most general bihermitian geometry requires
the introduction of semi-chiral $N=(2,2)$ superfields as well. This will appear elsewhere
\cite{wip}.

This paper is organized as follows. In the next section we briefly review
supersymmetric non-linear $\sigma $-models in $N=1$ boundary superspace.
Section 3 introduces $N=2$ boundary superspace together with the chiral
and twisted chiral superfields. In section 4 we determine the boundary
conditions which are allowed in the presence of chiral and twisted chiral
superfields. The results of section 4 are illustrated by several explicit
examples in section 5. The next section discusses duality transformations
which interchange chiral for twisted chiral fields and vice-versa. In
addition we also briefly discuss the duality between a pair consisting of
a chiral and a twisted chiral superfield and a semi-chiral multiplet. We
end with conclusions and an outlook. Our conventions are summarized in
the appendix.

\section{From $N=(1,1)$ to $N=1$}
In the absence of boundaries a non-linear $ \sigma $-model (with
$N\leq(1,1)$) on some $d$-dimensional target manifold $ {\cal M}$ is
characterized by a metric $g_{ab}(X)$ and a closed 3-form $T_{abc}(X)$
(the latter is known as the torsion, the Kalb-Ramond 3-form or the NS-NS
form) on $ {\cal M}$ where $X^a$ are local coordinates on ${\cal M}$ and
$a,b,c,...\in\{1,\cdots ,d\}$. The action in $N=(1,1)$ superspace is
simply\footnote{Our conventions are given in appendix A. Note that we
have rescaled the scalar fields with a factor $\sqrt{2 \pi \alpha '}$ in
order to make them dimensionless.},
\begin{eqnarray}
{\cal S}=8\int d^2 \sigma \, d^2 \theta \,D_+X^aD_-X^b\left(g_{ab}+b_{ab}\right),\label{an11}
\end{eqnarray}
where we used a locally defined 2-form potential $b_{ab}(X)=-b_{ba}(X)$
for the torsion,
\begin{eqnarray}
T_{abc}=- \frac{3}{2}\, \partial _{[a}b_{bc]}.
\end{eqnarray}

We consider a boundary at $ \sigma =0$ ( $ \sigma \geq 0$ ) and $ \theta
^+= \theta ^-$. This breaks the invariance under translations in both the
$ \sigma $ and the $  \theta' \equiv \theta ^+- \theta ^-$ direction thus
reducing the $N=(1,1)$ supersymmetry to an $N=1$ supersymmetry. We
introduce the derivatives,
\begin{eqnarray}
D\equiv D_++D_-,\qquad D'\equiv D_+-D_-,
\end{eqnarray}
which satisfy,
\begin{eqnarray}
D^2=D'{}^2=- \frac{i}{2} \partial _ \tau ,\qquad \{D,D'\}=-i\, \partial _ \sigma ,
\end{eqnarray}
and,
\begin{eqnarray}
D_+\,D_-=-\frac 1 2 \,D\,D' - \frac{i}{4}\,\partial _ \sigma.\label{ddsig}
\end{eqnarray}
The action,
\begin{eqnarray}
{\cal S}=-4\int d^2 \sigma \, d \theta
\,D'\left(D_+X^aD_-X^b\left(g_{ab}+b_{ab}\right)\right),\label{an1}
\end{eqnarray}
is manifestly invariant under the
surviving $N=1$ supersymmetry and -- because of eq.~(\ref{ddsig}) --
differs from the action in eq.~(\ref{an11}) by a boundary term \cite{Lindstrom:2002mc},
\cite{Koerber:2003ef}. Upon performing the
$D'$ derivative one gets the action in $N=1$ boundary superspace
previously obtained in \cite{Koerber:2003ef},
\begin{eqnarray}
{\cal S}&=&\int d^2 \sigma \, d \theta\Big(
i\,g_{ab}\,DX^a \partial _ \tau  X^b-2i\,g_{ab}\, \partial _ \sigma X^a D'X ^b+\, 2i\,b_{ab}\,
\partial _ \sigma X^a DX^b \nonumber\\
&&- 2 \,g_{ab}\, D'X ^a \nabla D'X ^b+2\,T_{abc} \,D'X ^aDX^bDX^c- \frac{2}{3} \,T_{abc}\,
D'X ^a D'X ^b D'X ^c\Big),\label{an12}
\end{eqnarray}
where,
\begin{eqnarray}
\nabla  D'X ^a\equiv D D'X ^a+ \left\{ {}^{\, a}_{bc} \right\}DX^b D'X ^c.
\end{eqnarray}
Both $X^a$ and $D'X^a$ are now independent $N=1$ superfields. Adding a boundary term $ {\cal S}_b$
to the action eq.~(\ref{an12}),
\begin{eqnarray}
{\cal S}_b=2i\,\int d \tau \, d \theta\,A_a\,DX^a, \label{Eva}
\end{eqnarray}
is equivalent to the modification $b_{ab}\rightarrow {\cal F}_{ab}
=b_{ab}+F_{ab} $ with $ F_{ab}= \partial _aA_b- \partial _b A_a$ in
eq.~(\ref{an12}).

Varying the action eq.~(\ref{an1})\footnote{Here one uses that $\int d^2
\sigma d \theta D'\,D_\pm=-(i/2)\,\int d \tau d \theta$. } or
eq.~(\ref{an12}) yields a boundary term,
\begin{eqnarray}
\delta {\cal S}\big| _{boundary}= -2i\,\int d \tau d \theta \,\delta X^a\left(
g_{ab} \,D'X ^b-\, b_{ab}\,DX^b\right),\label{var1}
\end{eqnarray}
which will only vanish upon imposing suitable boundary conditions. In order to do this
we introduce an almost product structure, a (1,1) tensor $R(X)^a{}_b$ \cite{Ooguri:1996ck}, 
\cite{Albertsson:2001dv},
\cite{Albertsson:2002qc},
\cite{Koerber:2003ef}, which satisfies,
\begin{eqnarray}
R^a{}_c\,R^c{}_b= \delta ^a_b,
\end{eqnarray}
and projection operators $ {\cal P}_\pm$,
\begin{eqnarray}
{\cal P}_\pm^a{}_b\equiv \frac{1}{2}\left( \delta ^a_b\pm R^a{}_b\right).
\end{eqnarray}
With this we impose Dirichlet boundary conditions,
\begin{eqnarray}
{\cal P}_-^a{}_b\, \delta X^b=0.\label{dbc1}
\end{eqnarray}
Using eq.~(\ref{dbc1}), one verifies that the boundary term eq.~(\ref{var1}) vanishes, provided one
imposes in addition the Neumann boundary conditions,
\begin{eqnarray}
{\cal P}_{+ba}\,D'X^b= {\cal P}_+^b{}_a\,b_{bc}\,DX^c.\label{nbc1}
\end{eqnarray}
If in addition we have that $R_{ab}=R_{ba}$ with $R_{ab}=g_{ac}R^c{}_b$,
then we can rewrite the Neumann boundary conditions as,
\begin{eqnarray}
 {\cal P}^a_+{}_bD' X^ b={\cal P}^a _+{}_cb^ c{}_d{\cal P}^d_+{}_b DX^b,
\end{eqnarray}
and ${\cal P}_+$ and ${\cal P}_-$ resp.~project onto Neumann and
Dirichlet directions resp. Note that as was discussed in
\cite{Sevrin:2007yn} this is not necessary.

Invariance of the
Dirichlet boundary conditions under what remains of the super-Poincar{\'e} transformations implies
that on the boundary,
\begin{eqnarray}
{\cal P}_-^a{}_b\,DX^b={\cal P}_-^a{}_b\, \partial _ \tau X^b=0,\label{ghghg}
\end{eqnarray}
hold as well. Using $D^2=-i/2\, \partial _ \tau $, we get from eq.~(\ref{ghghg}) the integrability
conditions\footnote{
Out of two $(1,1)$ tensors $R^a{}_b$ and $
S^a{}_b$, one constructs a $(1,2)$ tensor
${\cal N}[R,S]^a{}_{bc}$, the Nijenhuis tensor, as ${\cal N}[R,S]^a{}_{bc}=
R^a{}_dS^d{}_{[b,c]}+R^d{}_{[b}S^a{}_{c],d}+R\leftrightarrow S$.},
\begin{eqnarray}
0={\cal P}_+^d{}_{[b}{\cal P}_+^e{}_{c]} {\cal P}^a_+{}_{d,e}=
-\, \frac{1}{2}\, {\cal P}_-^a{}_e \,{\cal N}^e{}_{bc}[R,R].\label{ic11}
\end{eqnarray}
These conditions guarantee the existence of adapted coordinates $X^{\hat a}$, $\hat a
\in\{p+1,\cdots ,d\}$, with $p\leq d$ the rank of $ {\cal P_+}$ such that the Dirichlet boundary
conditions, eq.~(\ref{dbc1}) are simply given by,
\begin{eqnarray}
X^{\hat a}=\mbox{ constant}, \qquad \forall \,\hat a\in\{p+1,\cdots, d\}.\label{dbc9}
\end{eqnarray}
Writing the remainder of the coordinates as $X^{\check a}$, $\check a \in\{1,\cdots, p\}$, we get
the Neumann boundary conditions, eq.~(\ref{nbc1}), in our adapted coordinates,
\begin{eqnarray}
g_{\check a b}\,D'X^b=b_{\check a\check b}\,DX^{\check b},\label{nbc9}
\end{eqnarray}
where $b$ is summed from 1 to $d$ and we used that $DX^{\hat b}$ vanishes on the
boundary. Concluding, the action eq.~(\ref{an1}) together with the boundary
conditions eqs.~(\ref{dbc9}) and (\ref{nbc9}), describe open strings in the presence of a
D$p$-brane whose position is determined by eq.~(\ref{dbc9}).

\section{N=2 superspace}
\subsection{$N=(2,2)$ supersymmetry in the absence of boundaries}
Already in the absence of boundaries, promoting the $N=(1,1)$
supersymmetry of the action in eq.~(\ref{an11}) to an $N=(2,2)$
supersymmetry introduces additional structure. The most general extra
supersymmetry transformations -- consistent with dimensions and super
Poincar{\'e} symmetry -- are of the form,
\begin{eqnarray}
\delta X^a= \varepsilon ^+\,J_+^a{}_b(X)\,D_+X^b+\varepsilon ^-\,J_-^a{}_b(X)\,D_-X^b,\label{tr22}
\end{eqnarray}
which implies the introduction of two (1,1) tensors $J_+$ and $J_-$.
On-shell closure of the algebra requires both $J_+$ and $J_-$ to be complex structures,
\begin{eqnarray}
&&J_\pm^a{}_c\,J_\pm^c{}_b=- \delta ^a_b, \nonumber\\
&& N[J_\pm,J_\pm]^a{}_{bc}=0,
\end{eqnarray}
while invariance of the action necessitates that the metric is hermitian with respect
to {\em both} complex structures\footnote{This implies the existence
of two two-forms $ \omega^\pm _{ab}=-\omega^\pm _{ba}=
g_{ac}J^c_\pm{}_b$. In general they are not closed. Using
eq.~(\ref{covconst}), one shows that $ \omega ^\pm_{[ab,c]}=\mp 2
J_\pm^d{}_{[a}T_{bc]d}=\mp (2/3)J^d_\pm{}_a
J^e_\pm{}_bJ^f_\pm{}_cT_{def}$, where for the last step we used the
fact that the Nijenhuis tensors vanish.},
\begin{eqnarray}
J_\pm^c{}_a\,J_\pm^d{}_b\,g_{cd}=g_{ab}\,.\label{hermit}
\end{eqnarray}
and that both complex structures have to be covariantly constant,
\begin{eqnarray}
0=\nabla_c^\pm \,J_\pm^a{}_b\equiv
\partial _c\,J_\pm^a{}_b+\Gamma^a_{\pm dc}J_\pm^d{}_{b}-
\Gamma^d_{\pm bc}J_\pm^a{}_d\,,\label{covconst}
\end{eqnarray}
with the connections $\Gamma_\pm$ given by,
\begin{eqnarray}
\Gamma^a_{\pm bc}\equiv  \left\{ {}^{\, a}_{bc} \right\} \pm
T^a{}_{bc}\,.\label{cons}
\end{eqnarray}
The targetmanifold geometry $( {\cal M},g,J_\pm,T)$ consists of a
bihermitian manifold (the manifold has two complex structures for both of
which the metric is hermitian) and both complex structures are
covariantly constant with respect to different connections which are
given in eq.~(\ref{cons}). When the torsion vanishes, this type of
geometry reduces to the usual K{\"a}hler geometry.

An interesting observation is that all terms in the algebra which do not
close off-shell are proportional to the commutator of the complex
structures ${[}J_+,J_-{]}$ suggesting that extra auxiliary fields will be
needed in the direction of $\mbox{coker}{[}J_+,J_-{]}$. A detailed
analysis revealed the following picture (suggested in
\cite{Sevrin:1996jr}, \cite {Bogaerts:1999jc} and \cite{Ivanov:1994ec}
and shown to be correct in \cite{Lindstrom:2005zr}): writing
$\ker{[}J_+,J_-{]}=\ker(J_+-J_-)\oplus\ker(J_++J_-)$, one gets that
$\ker(J_+-J_-)$ and $\ker(J_++J_-)$ resp.~can be integrated to chiral and
twisted chiral multiplets resp.~\cite{Gates:1984nk}. Semi-chiral
multiplets \cite{Buscher:1987uw} are required for the description of
$\mbox{coker}{[}J_+,J_-{]}$. The Lagrange density is a real function of
these superfields. Metric, torsion and the complex structures can all be
expressed in terms of this function. When only chiral and twisted chiral
fields are present the relations are all linear while once semi-chiral
fields are present as well non-linearities appear. This clearly shows
that this geometry generalizes K{\"a}hler geometry: the whole local geometry
is encoded in a single real function which generalizes the K{\"a}hler
potential. As a consequence such geometries are often called generalized
K{\"a}hler geometries\footnote{This can be made very concrete in the
framework of Hitchin's generalized complex geometry, see {\em e.g.}
\cite{Gualtieri:2003dx}, \cite{Gualtieri:2007ng}, \cite{Lindstrom:2004hi}
and references therein.}.

In the present paper we will focus on chiral and twisted chiral
multiplets, {\em i.e.}~we assume that $J_+$ and $J_-$
commute\footnote{As already mentioned in the introduction we
relegate the study of the most general case -- which includes the
semi-chiral superfields -- to a forthcoming paper \cite{wip}.}.
These fields in $N=(2,2)$ superspace (once more we refer to the
appendix for conventions) satisfy the constraints $\hat D_\pm
X^a=J^a_\pm{}_b\,D_\pm X^b$ where $J_+$ and $J_-$ can be
simultaneously diagonalized. When the eigenvalues of $J_+$ and $J_-$
have the same (the opposite) sign we have chiral (twisted chiral)
superfields. Explicitly, we get that chiral superfields $ X^ \alpha
$, $ \alpha \in \{1,\cdots, m\}$, satisfy,
\begin{eqnarray}
\hat D_\pm X^ \alpha =+i\,D_\pm X^ \alpha ,\qquad
\hat D_\pm X^{ \bar \alpha }=-i\,D_\pm X^{ \bar \alpha }.\label{defsufi1}
\end{eqnarray}
Twisted chiral superfields $X^ \mu $, $ \mu \in \{1,\cdots, n\}$ satisfy,
\begin{eqnarray}
\hat D_\pm X^ \mu =\pm i\, D_\pm X^ \mu ,\qquad
\hat D_\pm X^{ \bar \mu }=\mp i\,D_\pm X^{ \bar \mu }.\label{defsufi2}
\end{eqnarray}
The most general action involving these superfields is given by,
\begin{eqnarray}
{\cal S}=4\,\int\,d^2 \sigma \,d^2\theta \,d^2 \hat \theta \, V(X, \bar X),
\end{eqnarray}
where the Lagrange density $V(X, \bar X)$ is an arbitrary real function of the chiral and twisted
chiral superfields. It is only defined modulo a generalized K{\"a}hler transformation,
\begin{eqnarray}
V\rightarrow V+F+ \bar F+ G+ \bar G,\label{genKahltrsf1}
\end{eqnarray}
with,
\begin{eqnarray}
\partial _{ \bar \alpha }F= \partial _{ \bar \mu }F=0,\qquad
\partial _{ \bar \alpha }G= \partial _ \mu G=0.\label{genKahltrsf2}
\end{eqnarray}
Passing to $N=(1,1)$ superspace and comparing the result to eq.~(\ref{an11}),
allows one to identify the metric and the torsion potential\footnote{Indices from the beginning of
the Greek alphabet, $ \alpha $, $ \beta $, $ \gamma $, ... denote chiral fields while indices from
the middle of the alphabet, $ \mu $, $ \nu $, $ \rho $, ... denote twisted chiral fields.},
\begin{eqnarray}
&&g_{ \alpha \bar \beta }=+V_{ \alpha \bar \beta },\qquad g_{ \mu \bar \nu }= -
V_{ \mu \bar \nu },\nonumber\\
&&b_{ \alpha \bar \nu }=-V_{ \alpha \bar \nu }, \qquad b_{ \mu \bar \beta  }=+V_{ \mu \bar \beta  },
\label{KRform}
\end{eqnarray}
where all other components of $g$ and $b$ vanish. When writing $V_{
\alpha \bar \beta }$, we mean $ \partial _ \alpha \partial_ { \bar \beta
}V$ etc. Let us end with two remarks. Interchanging the chiral with the
twisted chiral superfields and vice-versa while sending $V\rightarrow
-V$, leaves the bulk geometry unchanged. Finally, when only one type of
superfield is present, the geometry is K{\"a}hler.

\subsection{From $N=(2,2)$ to $N=2$}
We introduce a boundary in $N=(2,2)$ superspace which breaks half of the
supersymmetries, reducing $N=(2,2)$ to $N=2$. We have either B-type
boundary conditions where the boundary is given by $ \theta '\equiv
(\theta ^+- \theta ^-)/2=0$ and $ \hat \theta '\equiv(\hat \theta ^+-\hat
\theta ^-)/2=0$ or A-type boundary conditions where the boundary is given
by $ \theta '\equiv (\theta ^+- \theta ^-)/2=0$ and $ \hat \theta
'\equiv(\hat \theta ^++\hat \theta ^-)/2=0$. Throughout this paper we
will always use B-type boundary conditions as switching to A-type
boundary conditions merely amounts to interchanging chiral fields for
twisted chiral fields and vice-versa \cite{Sevrin:2007yn}.

We define the derivatives,
\begin{eqnarray}
&&D\equiv D_++D_-,\qquad \hat D\equiv \hat D_++ \hat D_-, \nonumber\\
&& D'\equiv D_+-D_-,\qquad \hat D'\equiv \hat D_+- \hat D_-,\label{Aders}
\end{eqnarray}
where unaccented derivatives refer to translations in the invariant directions.
They satisfy,
\begin{eqnarray}
&&D^2=\hat D^2=D'{}^2=\hat D'{}^2=- \frac{i}{2} \partial _ \tau , \nonumber\\
&&\{D,D'\}=\{\hat D, \hat D'\}=-i \partial _ \sigma ,
\end{eqnarray}
with all other anti-commutators zero.

Let us now turn to the superfields. In the bulk we had chiral and twisted chiral
superfields. From
eqs.~(\ref{defsufi1}) and (\ref{Aders}) we get for the chiral fields,
\begin{eqnarray}
&&\hat D X^ \alpha =+i DX^ \alpha ,\quad
\hat D X^ {\bar \alpha} =-i DX^{\bar \alpha}  \nonumber\\
&&\hat D' X^ \alpha =+i D'X^ \alpha ,\quad
\hat D' X^ {\bar \alpha} =-i D'X^{\bar \alpha} ,\label{bcf}
\end{eqnarray}
where $ \alpha,\ \bar \alpha  \in\{1,\cdots, m\}$. This can also be
written as\footnote{For our conventions we refer once more to the
appendix.},
\begin{eqnarray}
\bar \ID X^ \alpha= \bar \ID'X^ \alpha =
\ID X^{ \bar \alpha }=\ID' X^{ \bar \alpha }=0.
\end{eqnarray}
Passing from $N=(2,2)$ -- parametrized by the Grassmann coordinates $ \theta $, $ \hat \theta $,
$ \theta '$ and $ \hat \theta '$ -- to $N=2$ superspace --
parametrized by $ \theta $ and $ \hat \theta $ -- we get $X^ \alpha $, $X^{ \bar \alpha }$,
$D'X^ \alpha $ and $D'X^{ \bar \alpha }$ as $N=2$ superfields which satisfy the constraints,
\begin{eqnarray}
&&\hat D X^ \alpha =+i\,D X^ \alpha ,\qquad
\hat D X^{ \bar \alpha }=-i\, D X^{ \bar \alpha }, \nonumber\\
&&\hat D \,D'X^ \alpha =+i\,D \,D'X^ \alpha- \partial _ \sigma X^ \alpha  ,\qquad
\hat D\,D' X^{ \bar \alpha }=-i\, D\,D' X^{ \bar \alpha }+ \partial _ \sigma X^{ \bar \alpha }.
\label{n2csf}
\end{eqnarray}

For twisted chiral superfields we get instead, when combining
eqs.~(\ref{defsufi2}) and (\ref{Aders}),
\begin{eqnarray}
&&\hat D X^ \mu =+i D'X^ \mu ,\quad
\hat D X^ {\bar \mu} =-i D'X^{\bar \mu},  \nonumber\\
&&\hat D' X^ \mu =+i DX^ \mu ,\quad
\hat D' X^ {\bar \mu} =-i DX^{\bar \mu} ,\label{btcf}
\end{eqnarray}
with $ \mu ,\ \bar \mu \in\{1,\cdots, n\}$. For further convenience we can also write this as,
\begin{eqnarray}
&&\ID'X^ \mu =\ID X^ \mu ,\qquad \bar \ID'X^ \mu =- \bar \ID X^ \mu , \nonumber\\
&&\ID'X^{ \bar \mu } =-\ID X^ { \bar \mu } ,\qquad \bar \ID'X^{ \bar \mu }
= \bar \ID X^ { \bar \mu }.
\end{eqnarray}
Passing again
from $N=(2,2)$ to $N=2$ superspace, we now get $X^ \mu  $, $X^{ \bar \mu  }$,
$D'X^ \mu  $ and $D'X^{ \bar \mu  }$ as $N=2$ superfields satisfying the constraints,
\begin{eqnarray}
&&\hat D X^{ \mu }=+i\, D'X^{ \mu },\quad
\hat D X^{ \bar  \mu }=-i\, D'X^{ \bar  \mu }, \nonumber\\
&& \hat D \,D'X^{ \mu }=- \frac{1}{2}\dot X^{ \mu },\quad
\hat D\,D' X^{ \bar  \mu }=+ \frac{1}{2}\dot X^{ \bar  \mu }.\label{n2tcsf}
\end{eqnarray}
Note that in $N=2$ boundary superspace, the twisted chiral superfields
$X^ \mu $ and $X^{ \bar \mu }$ are unconstrained superfields -- the
fermionic fields $D'X$ are nothing else but the image of these fields
under the second supersymmetry -- while the chiral fields can be viewed
as a 1-$d$ analogue of chiral fields.

Once more one immediately verifies that the difference between the
fermionic measure $D_+D_- \hat D_+\hat D_-$ and $D \hat D D' \hat D'$ is
just a boundary term. So the most general $N=2$ invariant action which
reduces to the usual action away from the boundary that we can write down
is,
\begin{eqnarray}
{\cal S}&=&-\int d^2 \sigma\, d \theta d \hat \theta\, D' \hat D'\, V(X, \bar X)+
i\,\int d \tau \,d \theta d \hat \theta \,W( X, \bar X) \nonumber\\
&=&-\int d^2 \sigma\, d^2 \theta\, D' \hat D'\, V(X, \bar X)+
i\,\int d \tau \,d^2 \theta  \,W( X, \bar X),\label{bsfa}
\end{eqnarray}
with $V(X, \bar X)$ and $W( X , \bar X)$ real functions of the chiral and
the twisted chiral superfields. While the generalized K{\"a}hler potential
$V$ is arbitrary, this is not so for the boundary potential $W$. Whenever
$W$ is a function of the twisted chiral fields as well, its form will be
(partially) determined by the boundary conditions as we will see later
on.

\section{Non-linear $ \sigma $-models}
\subsection{The action}
We start with a set of chiral superfields $X^ \alpha $, $ \alpha
\in\{1,\cdots m\}$, and a set of twisted chiral superfields $X^ \mu $, $
\mu \in\{1,\cdots , n\}$. The action is given by eq.~(\ref{bsfa}).
Working out the $ \hat D'$ and $D'$ derivatives we obtain the action in
$N=2$ superspace\footnote{When comparing this to the K{\"a}hler case
discussed in \cite{Sevrin:2007yn}, note that when no twisted chiral
fields are present, $\int d^2 \sigma d^2 \theta V_ \alpha  \partial _
\sigma X^ \alpha = -2i\int d^2 \sigma d^2 \theta V_{ \bar  \alpha \beta
}DX^{ \bar \alpha }D'X^ \beta $ holds.},
\begin{eqnarray}
{\cal S}&=&\int d^2 \sigma\, d^2 \theta\,
\big\{
+2i\,V_{ \alpha \bar \beta }D'X^ \alpha D'X^{ \bar \beta }
-2i\,V_{ \alpha \bar \beta }DX^ \alpha DX^{ \bar \beta }
+2i\,V_{ \mu \bar \nu }DX^ \mu D'X^{ \bar \nu }\nonumber\\
&&\qquad
+2i\,V_{ \mu \bar \nu }D'X^ \mu DX^{ \bar \nu }
-V_ \mu \partial _ \sigma X^ \mu +V_{ \bar \mu } \partial _ \sigma X^{ \bar \mu }
-V_ \alpha  \partial _ \sigma X^ \alpha +V_{ \bar \alpha } \partial _ \sigma X^{ \bar \alpha }
\nonumber\\
&&\qquad
+2i\,V_{ \alpha \bar \beta }D'X^ \alpha DX^{ \bar \beta }
+2i\,V_{ \alpha \bar \beta }DX^ \alpha D'X^{ \bar \beta }
+2i\,V_{ \alpha \bar \nu }D'X^ \alpha D'X^{ \bar \nu }\nonumber\\
&&\qquad
+2i\,V_{ \mu \bar \beta }D'X^ \mu D'X^{ \bar \beta }
+2i\,V_{ \mu \bar \beta }DX^ \mu D'X^{ \bar \beta }
+2i\,V_{ \alpha \bar \nu }D'X^ \alpha DX^{ \bar \nu }
\big\} \nonumber\\
&&\qquad\qquad +
i\,\int d \tau \,d^2 \theta  \,W( X, \bar X).\label{bsfaa}
\end{eqnarray}
When reducing the action to $N=1$ superspace, one recovers the action in eq.~(\ref{an12})
with metric and torsion given by eq.~(\ref{KRform}). However the resulting action has a
boundary term as well,
\begin{eqnarray}
{\cal S}_{boundary}&=&-i\,\int d \tau \,d \theta \,
\Big(\big(V_ \alpha -i W_ \alpha \big)DX^ \alpha +
\big(V_{ \bar \alpha }+iW_{ \bar \alpha }\big)DX^{ \bar \alpha }+ \nonumber\\
&&\qquad\big(V_ \mu -i W_ \mu \big)D'X^ \mu +
\big(V_{ \bar \mu }+iW_{ \bar \mu }\big)D'X^{ \bar \mu }\Big).\label{Eva1}
\end{eqnarray}
Note that this boundary term -- because of the presence of $D'X$ terms --
does not have the standard form (compare to eq.~(\ref{Eva})). A judicious
choice of boundary conditions will allow us to reduce it to the form in
eq.~(\ref{Eva}).

The action is still invariant under the generalized K{\"a}hler transformations,
eqs.~(\ref{genKahltrsf1}) and (\ref{genKahltrsf2}), provided the boundary potential $W$
transforms as well,
\begin{eqnarray}
W\rightarrow W-i\,\big(F- \bar F\big)+i\,\big(G- \bar G\big).\label{genKahlbd}
\end{eqnarray}
Invariance under generalized K{\"a}hler transformations is essential for the
global consistency of the models. Let us illustrate this with a simple
example -- more and less trivial examples will follow later in the paper
-- of a D1-brane on a two-torus $T^2$. The torus is characterized by its
modulus $\tau$ which takes its value in the upper half-plane $\IH$. We
parametrize the torus by a twisted chiral field $w=(x+\tau y)/\sqrt{2}$
with $x,\,y\in\IR$, such that the metric is simply $g_{w\bar w}=1$. The
periodicity condition is,
\begin{eqnarray}
 w\simeq w+ \frac{1}{\sqrt{2}}\big(n_1+n_2\,\tau \big),\label{tn1}
\end{eqnarray}
with $n_1,\,n_2\in\IZ$. We impose the Dirichlet boundary condition,
\begin{eqnarray}
 \big(1+m\,\bar \tau \big)\,w=\big(1+m\,\tau \big)\,\bar w,\label{tn2}
\end{eqnarray}
with $m\in\IZ$. Because of eq.~(\ref{n2tcsf}) this implies a Neumann
boundary condition as well,
\begin{eqnarray}
 \big(1+m\,\bar \tau \big)\,D'w+\big(1+m\,\tau \big)\,D'\bar w=0,
\end{eqnarray}
and we end up with a D1-brane winding once in the $x$ direction and $m$
times in the $y$ direction. The K{\"a}hler potential is $V=-w \bar w$ and
with the boundary condition eq.~(\ref{tn2}) one finds\footnote{One can
use the results in \cite{Sevrin:2007yn} or require that eq.~(\ref{bsfab})
vanishes.} that the boundary potential vanishes, $W=0$. Because of the
presence of a D1-brane, the invariance eq.~(\ref{tn1}) is partially
broken and we get from eq.~(\ref{tn2}) that,
\begin{eqnarray}
 n_2=m\,n_1,\label{tn3}
\end{eqnarray}
should hold. Under eqs.~(\ref{tn1}) and (\ref{tn3}), the K{\"a}hler
potential transforms as,
\begin{eqnarray}
 V\rightarrow V-\frac{n_1}{\sqrt{2}}\big(1+m\,\bar \tau \big)w-
 \frac{n_1}{\sqrt{2}}\big(1+m\, \tau \big)\bar w.
\end{eqnarray}
Making a K{\"a}hler transformation restores the invariance but generates --
because of eq.~(\ref{genKahlbd}) -- a boundary potential,
\begin{eqnarray}
 W=0\rightarrow W= -\frac{i\,n_1}{\sqrt{2}}\left(
\big(1+m\,\bar \tau \big)\,w-\big(1+m\,\tau \big)\,\bar w \right),
\end{eqnarray}
which vanishes because of the boundary condition eq.~(\ref{tn2}). So the
description is indeed globally consistent.

Finally note that the action eq.~(\ref{bsfa}) is also invariant under,
\begin{eqnarray}
W\rightarrow W+H+ \bar H,\label{awq1}
\end{eqnarray}
where,
\begin{eqnarray}
\partial _{ \bar \alpha }H= \partial _ \mu H= \partial _{ \bar \mu }H=0,\qquad
\partial _{\alpha } \bar H= \partial _ \mu \bar H= \partial _{ \bar \mu } \bar H=0.\label{awq2}
\end{eqnarray}
We will often tacitly use the fact that the boundary potential is only
defined modulo an additive contribution of a holomorphic (and its complex
conjugate) function of the boundary chiral fields.

When varying the action, one needs to take into account that the superfields are constrained.
Besides $X^ \mu $ and $X^{ \bar \mu }$ we introduce unconstrained superfields
$ \Lambda ^ \alpha $, $ \Lambda ^{ \bar \alpha }$, $M^ \alpha $ and $M^{ \bar \alpha }$ and
solve the constraints by,
\begin{eqnarray}
&&X^ \alpha = \bar \ID \Lambda ^ \alpha ,\qquad
X^{ \bar \alpha }=\ID \Lambda ^{ \bar \alpha }, \nonumber\\
&&D'X^ \alpha = \bar \ID M^{ \alpha }- \partial _ \sigma \Lambda ^{ \alpha },\qquad
D'X^{ \bar \alpha } = \ID M^{ \bar \alpha }+ \partial _ \sigma \Lambda ^{ \bar \alpha }
\nonumber\\
&&D'X^ \mu =-i\, \hat DX^ \mu ,\qquad
D'X^{ \bar \mu }=+i \hat DX^{ \bar \mu }.
\end{eqnarray}
Varying the unconstrained fields in the action eq.~(\ref{bsfaa}) yields the usual equations of
motion with metric and Kalb-Ramond two-form given by eq.~(\ref{KRform}) and a boundary
term given by,
\begin{eqnarray}
\delta {\cal S}\Big|_{boundary}&=&\int d \tau \,d^2 \theta  \,
\Big\{
\delta \Lambda ^{ \alpha } \big(
\bar \ID'V_ { \alpha }+i\, \bar \ID W_{ \alpha }
\big)-
\delta \Lambda ^{ \bar \alpha } \big(
 \ID'V_ { \bar \alpha }-i \, \ID W_{ \bar \alpha }\big)- \nonumber\\
&&\qquad
 \delta X^{ \mu } \big(V_{ \mu }-i W_{ \mu }\big)+
 \delta X^{ \bar \mu } \big(V_{ \bar \mu }+i W_{ \bar \mu }\big)
\Big\}.\label{bsfab}
\end{eqnarray}
This should vanish by imposing appropriate boundary conditions on the fields.

\subsection{Boundary conditions} \label{subs:bc}
\subsubsection{General considerations}
We will impose Dirichlet boundary conditions using an almost product
structure as was introduced in section 2. We start with the unconstrained
superfields $ \Lambda $. The most general Dirichlet boundary conditions
which are consistent with the dimensions of the fields involved are given
by,
\begin{eqnarray}
\delta \Lambda ^ \alpha =R(X)^ \alpha {}_{ \beta } \delta \Lambda ^{ \beta  }+
R(X)^ \alpha {}_{ \bar \beta } \delta \Lambda ^{ \bar \beta },\label{bc1}
\end{eqnarray}
which already implies that,
\begin{eqnarray}
R^ \alpha {}_{ \mu }= R^ \alpha {}_{ \bar \mu }=0.\label{cbc1}
\end{eqnarray}
As $ \bar \ID \delta \Lambda ^{ \bar \beta }$ should not appear in the boundary condition
for $X^ \alpha $ we necessarily have that,
\begin{eqnarray}
R^ \alpha {}_{ \bar \beta }=0.\label{cbc2}
\end{eqnarray}
Eq.~(\ref{bc1}) implies,
\begin{eqnarray}
\delta X^ \alpha =R^ \alpha {}_ \beta \, \delta X^ \beta ,\label{bc2}
\end{eqnarray}
if,
\begin{eqnarray}
&&R^ \alpha {}_{ \delta , \bar \epsilon }{\cal P}_+^ \delta {}_ \beta
{\cal P}_+^{ \bar \epsilon }{}_{ \bar \gamma }+
R^ \alpha {}_{ \delta , \mu }{\cal P}_+^ \delta {}_ \beta {\cal P}_+^{ \mu }{}_{ \bar \gamma }
+R^ \alpha {}_{ \delta , \bar \mu }{\cal P}_+^ \delta {}_ \beta {\cal P}_+^{ \bar \mu }{}_{ \bar 
\gamma }=0,
\nonumber\\
&&R^ \alpha {}_{ \delta , \nu }{\cal P}_+^ \delta {}_ \beta {\cal P}_+^{ \nu }{}_ \mu +
R^ \alpha {}_{ \delta , \bar \nu }
{\cal P}_+^ \delta {}_ \beta {\cal P}_+^{ \bar \nu }{}_ \mu=0, \nonumber\\
&&R^ \alpha {}_{ \delta , \nu }{\cal P}_+^ \delta {}_ \beta
{\cal P}_+^{ \nu }{}_ { \bar \mu } +
R^ \alpha {}_{ \delta , \bar \nu }{\cal P}_+^ \delta {}_ \beta
{\cal P}_+^{ \bar \nu }{}_ { \bar \mu }=0,
\label{bic1}
\end{eqnarray}
holds. Furthermore, eq.~(\ref{bc2}) implies $D X^ \alpha =R^ \alpha {}_ \beta \, D X^ \beta$,
$ \hat D X^ \alpha =R^ \alpha {}_ \beta \, \hat D X^ \beta$ and
$\dot X^ \alpha =R^ \alpha {}_ \beta \, \dot X^ \beta$ as well. Consistency of this
with $D^2= \hat D^2=-(i/2) \partial _ \tau $ results in the integrability condition,
\begin{eqnarray}
R^ \alpha {}_{ \delta , \epsilon } \,{\cal P_+}^ \delta {}_{[ \beta }\,
{\cal P_+}^ \epsilon  {}_{ \gamma ] }=0.\label{bic2}
\end{eqnarray}
Eqs.~(\ref{bic1}) and (\ref{bic2}) together form the integrability
conditions (\ref{ic11}) for $a=\alpha$ and thus guarantee the existence
of a holomorphic coordinate transformation which brings us to coordinates
$X^{\hat \alpha } $, $ \hat\alpha \in\{k+1,\cdots m\}$
 where $2k$ is the rank of $ {\cal P}_+$ (in the chiral directions), such that part of the Dirichlet
boundary conditions are given by,
\begin{eqnarray}
X^{\hat \alpha }=\mbox{ constant},\label{dcl1}
\end{eqnarray}
with $X^{\hat \alpha }$ chiral. We will denote the remainder of the
chiral coordinates by $X^{\tilde \alpha }$, $\tilde \alpha \in\{1,\cdots,
k\}$. In these coordinates we have that,
\begin{eqnarray}
R^{ \hat \alpha }{}_{ \tilde \beta }=0,\qquad
R^{ \tilde \alpha }{}_{ \tilde \beta }= \delta ^{ \tilde \alpha }_{ \tilde \beta }, \label{cac1}
\end{eqnarray}
and,
\begin{eqnarray}
R^{ \hat \alpha }{}_{ \hat \gamma }R^{ \hat \gamma }{}_{ \hat \beta }=
\delta ^{ \hat \alpha }_{ \hat \beta },\qquad
R^{ \tilde \alpha }{}_{ \hat \gamma }R^{ \hat \gamma }{}_{ \hat \beta }=
-R^{ \tilde \alpha }{}_{ \hat \beta }.
\end{eqnarray}
For the time being however, we only require the chiral fields to obey (\ref{bc2}), without going
to these adapted coordinates.

We now turn to the Dirichlet boundary conditions for the twisted chiral
superfields. The most general expression we can write down is,
\begin{eqnarray}
\delta X^ \mu = R^ \mu {}_ \nu \,\delta X^ \nu +R^ \mu {}_{ \bar \nu }\, \delta X^{ \bar \nu }+
R^ \mu {}_ \beta \, \delta X^\beta +
R^ \mu {}_{\bar\beta}\, \delta X^{\bar\beta}.\label{dcl2}
\end{eqnarray}
Using eqs.~(\ref{n2tcsf}) and (\ref{n2csf}), we get from this,\footnote{Note that the contributions
from $DX^{\hat\alpha}$ (and its complex conjugate) to this expression actually vanish because
of (\ref{bc2}), but the ones from $D'X^{\hat\alpha}$ (and its complex conjugate )do not.}
\begin{eqnarray}
\big({\cal P}_+D'X\big)^ \mu =
R^ \mu {}_ \nu \,D'X^ \nu + \frac{1}{2}R^ \mu {}_\beta \,\big(D'X^\beta
+DX^\beta \big)+\frac{1}{2}R^ \mu {}_{\bar\beta }\,
\big(D'X^{\bar\beta }
-DX^{\bar\beta }\big). \label{gnc1}
\end{eqnarray}
This is consistent with $ {\cal P}_+^2= {\cal P}_+$ if,
\begin{eqnarray}
&&R^ \mu {}_ \rho R^ \rho {}_ \nu =R^ \mu {}_ \nu , \nonumber\\
&&R^\mu {}_\rho R^\rho {}_{\bar \nu }=R^ \mu {}_{ \bar \rho }R^{ \bar \rho }{}_{ \bar \nu }=0, \label{conr1}
\end{eqnarray}
and
\begin{eqnarray}
R^ \mu {}_ \beta &=& R^ \mu {}_ \nu R^ \nu {}_ \beta - R^ \mu {}_{ \bar \nu }R^{ \bar \nu }{}_\beta , \nonumber\\
R^ \mu {}_{\bar \beta} &=& R^ \mu {}_ \nu R^ \nu {}_{\bar \beta} - R^ \mu {}_{ \bar \nu }
R^{ \bar \nu }{}_{\bar\beta}. \label{conr2}
\end{eqnarray}
Using the defining property of an almost product structure --
$R^a{}_cR^c{}_b=R^a{}_b$ -- and eqs.~(\ref{cbc1}) and (\ref{conr1}), one
finds that both $\pi_+^ \mu {}_ \nu \equiv R^ \mu {}_ \nu $ and $ \pi _-^
\mu {}_ \nu\equiv \delta ^ \mu _ \nu - R^ \mu {}_ \nu =R^ \mu {}_{ \bar
\rho } R^{ \bar \rho }{}_ \nu  $ are projection operators mapping
$T^{(1,0)}_{ {\cal M}}$ to $T^{(1,0)}_{ {\cal M}}$. In terms of these
projection operators eqs.~(\ref{conr2}) can be rewritten more
suggestively as,
\begin{eqnarray}
R^ \mu {}_ \beta &=& \left( \pi _-^ \mu {}_ \nu - \pi _+^ \mu {}_ \nu \right) R^ \nu {}_ \alpha
R^ \alpha {}_ \beta ,\nonumber\\
R^ \mu {}_ {\bar\beta} &=& \left( \pi _-^ \mu {}_ \nu - \pi _+^ \mu {}_ \nu \right) R^ \nu {}_ {\bar\alpha}
R^ {\bar\alpha} {}_ {\bar\beta}.
\end{eqnarray}
In the $\pi_-$ directions these relations are trivially satisfied in chiral directions along the brane,
as follows from (\ref{bc2}) or (\ref{cac1}), and hence have no consequences for the Dirichlet
conditions (\ref{dcl2}). In the $\pi_+$ directions, however, they imply,
\begin{eqnarray}
 \pi _+^ \mu {}_ \nu R^ \nu {}_ \beta \delta X^\beta =  \pi _+^ \mu {}_ \nu R^ \nu {}_ {\bar\beta}
 \delta X^{\bar\beta} = 0,
\end{eqnarray}
or
\begin{eqnarray}
 \pi _+^ \mu {}_ \nu R^ \nu {}_ {\tilde\beta}  =  \pi _+^ \mu {}_ \nu R^ \nu {}_ {\bar{\tilde\beta}}  = 0.
\end{eqnarray}
This implies that there are no Dirichlet conditions in the $\pi_+$ directions, as can be seen by
acting with $\pi_+$ on both sides of eq.~(\ref{dcl2}).
This is made manifest by writing the Dirichlet boundary conditions
eq.~(\ref{dcl2}) as
\begin{eqnarray}
\pi _-^ \mu {}_ \nu \delta X^ \nu =
R^ \mu {}_{ \bar \nu }\, \delta X^{ \bar \nu }+
R^ \mu {}_ {\tilde \beta }\, \delta X^{ \tilde \beta }+
R^ \mu {}_{\bar{\tilde \beta }}\, \delta X^{\bar{\tilde \beta }}.\label{dcl3}
\end{eqnarray}
The corresponding Neumann boundary conditions are then,
\begin{eqnarray}
\pi _-^ \mu {}_ \nu D'X^ \nu =
-R^ \mu {}_{ \bar \nu }\, D' X^{ \bar \nu }+
R^ \mu {}_ {\tilde \beta }\, D X^{ \tilde \beta }-
R^ \mu {}_{\bar{\tilde \beta }}\, D X^{\bar{\tilde \beta }}.\label{dcl4}
\end{eqnarray}
We need to impose separate Neumann boundary conditions on $\pi _+^ \mu
{}_ \nu D'X^ \nu $. In the $\pi_+$ directions (\ref{gnc1}) can be written
as,
\begin{eqnarray}
 \pi _+^ \mu {}_ \nu\big({\cal P}_+D'X\big)^ \nu =
R^ \mu {}_ \nu \left( D'X^ \nu + \frac{1}{2}R^ \nu {}_{\hat\beta} \,D'X^{\hat\beta}
+ \frac{1}{2}R^ \nu {}_{\bar{\hat\beta} }\,
D'X^{\bar{\hat\beta} } \right), \label{gnc2}
\end{eqnarray}
which, comparing to {\em e.g.} eq.~(\ref{nbc9}), shows that there is a
non-degenerate $U(1)$ field strength in the $\pi_+$ directions. A similar
expression in the $\pi_-$ directions
\begin{eqnarray}
 \pi _-^ \mu {}_ \nu\big({\cal P}_+D'X\big)^ \nu = \frac{1}{2}
\pi_-^ \mu {}_ \nu \left( R^ \nu {}_\beta \,\big(D'X^\beta
+DX^\beta \big)+ R^ \nu {}_{\bar\beta }\,
\big(D'X^{\bar\beta }
-DX^{\bar\beta }\big) \right), \label{gnc3}
\end{eqnarray}
indicates that we can expect a field strength in these directions as well, as long as $R^ \mu {}_\beta$
and $R^ \mu {}_{\bar\beta}$ are non-vanishing.

\subsubsection{Detailed analysis}

For simplicity, let us first examine the extremal cases $\pi_- = 1$ and $\pi_+ = 1$. When $\pi_- = 1$,
we get an equal amount of Dirichlet and Neumann conditions on the twisted chiral fields. To make
(\ref{bsfab}) vanish, we start by setting,
\begin{eqnarray}
(V-iW)_\mu R^ \mu {}_{ \bar \nu } = (V+iW)_{\bar\nu} .
\end{eqnarray}
Using this, one rewrites the Dirichlet conditions eq.~(\ref{dcl3}) as,
\begin{eqnarray}
(V-iW)_\mu \left( \delta X^ \mu - R^\mu{}_\beta \delta X^\beta \right) = (V+iW)_{\bar\mu}
\left( \delta X^{\bar\mu} - R^{ \bar \mu }{}_{\bar\beta} \delta X^{\bar\beta}  \right), \label{pimind}
\end{eqnarray}
which also implies $R^ \mu {}_\beta = - R^ \mu {}_{ \bar \nu }R^{ \bar \nu }{}_\beta$ or,
\begin{eqnarray}
(V-iW)_\mu R^ \mu {}_\beta = -(V+iW)_{\bar\mu} R^{ \bar \mu }{}_\beta .
\end{eqnarray}
Compatibility of eq.~(\ref{pimind}) with $D^2 \propto \partial_\tau$
results in the extra condition
\begin{eqnarray}
R_{\mu\nu} = R_{\nu\mu},
\end{eqnarray}
along with the conditions which insure integrability of ${\cal P}_+$,
eq.~(\ref{ic11}), already known from the $N=1$ superspace analysis. The
Dirichlet conditions (\ref{pimind}) again automatically imply the Neumann
conditions
\begin{eqnarray}
(V-iW)_\mu \left( D' X^ \mu - R^\mu{}_\beta D X^\beta \right) + (V+iW)_{\bar\mu}
\left( D' X^{\bar\mu} - R^{ \bar \mu }{}_{\bar\beta} D X^{\bar\beta}  \right) = 0. \label{piminn1}
\end{eqnarray}
Using these equations to simplify the $N=1$ boundary term (\ref{Eva1}) and comparing the
result with (\ref{Eva}) yields a $U(1)$ connection with components
\begin{eqnarray}
A_{\tilde\alpha} &=& - \frac 1 2 \left[ (V-iW)_{\tilde\alpha} + (V-iW)_\mu R^\mu{}_{\tilde\alpha}
\right], \nonumber\\
A_{\bar{\tilde\alpha}} &=& - \frac 1 2 \left[ (V+iW)_{\bar{\tilde\alpha}} + (V+iW)_{\bar\mu}
R^{\bar\mu}{}_{\bar{\tilde\alpha}} \right] ,\label{piminA}\\
A_\mu &=& A_{\bar\mu} = 0, \nonumber
\end{eqnarray}
up to gauge transformations.
Using (\ref{pimind}) in (\ref{bsfab}) yields,
\begin{eqnarray}
\delta {\cal S}\Big|_{boundary}&=&\int d \tau \,d^2 \theta  \,
\Big\{
\delta \Lambda ^{ \alpha } \Big[
\bar \ID'V_ { \alpha }- \bar \ID \big((V-iW)_\mu R^ \mu {}_\alpha - i\,W_{ \alpha } \big)
\Big]\nonumber\\
&&\qquad\qquad-\delta \Lambda ^{ \bar \alpha } \Big[
 \ID'V_ { \bar \alpha }- \ID \big((V+iW)_{\bar\mu} R^{ \bar \mu }{}_{\bar\alpha} + i\,
 W_{ \bar \alpha} \big)\Big]\Big\}.\label{bsfab2}
\end{eqnarray}
This leads to the Neumann conditions
\begin{eqnarray}
\bar \ID'V_ { \tilde\alpha } &=& \bar \ID \big((V-iW)_\mu R^ \mu {}_{\tilde\alpha} - i\,
W_{ \tilde\alpha } \big) ,\nonumber\\
 \ID'V_ { \bar{\tilde \alpha} } &=& \ID \big((V+iW)_{\bar\mu} R^{ \bar \mu }{}_{\bar{\tilde\alpha}}
 + i\, W_{ \bar{\tilde \alpha}} \big). \label{piminn2}
\end{eqnarray}
Eqs.~(\ref{piminn2}) together with (\ref{dcl1}), (\ref{pimind}) and
(\ref{piminn1}), describe a ($2k+n$)-dimensional brane, where $k$ is the
number of chiral fields along the brane and $n$ is the number of twisted
chiral fields. In principle, it should be possible to demonstrate that
(\ref{piminn1}) and (\ref{piminn2}) are precisely of the general form
(\ref{nbc9}). This however requires the introduction of worldvolume
coordinates (which solve eqs.~(\ref{pimind})). We will not show this in
full generality here, but will illustrate this point for some examples in
section \ref{section:ex}.

Let us now turn to the case $\pi_+ = 1$. As mentioned before, in this case all
boundary conditions on the twisted chiral fields are necessarily Neumann. Because of the
presence of a $U(1)$ field strength $F$ -- as implied by (\ref{gnc2}) -- we expect a
condition of the form
\begin{eqnarray}
D'X^k = F^k{}_l DX^l + (F^k{}_a + b^k{}_a) D X^a. \label{piplusn}
\end{eqnarray}
Here, Latin indices from the beginning of the alphabet indicate (both
holomorphic and anti-holomorphic) chiral directions and Latin indices
from the middle of the alphabet indicate twisted chiral directions. Note
that we have taken into account the non-trivial b-field background
(\ref{KRform}). Writing the $N=2$ relations as $\hat{D}X^a=J^a{}_bDX^b$
and $\hat{D}X^m=J^m{}_nD'X^n$, we find that eq.~(\ref{piplusn})
implies\footnote{Notice that components $L^k{}_{\hat a}$ will always be
zero, since there is no magnetic field in these directions. This is
however implicit in all subsequent formulae, because in the end $DX^{\hat
a}$ will be zero as well by virtue of the Dirichlet conditions in the
chiral directions.},
\begin{eqnarray}
\hat D X^k = K^k{}_l DX^l + L^k{}_a D X^a, \label{boc}
\end{eqnarray}
where $K^k{}_l = J^k{}_m F^m{}_l$ and $L^k{}_a = J^k{}_m (F^m{}_a + b^m{}_a)$. This means that on
the boundary the twisted chiral fields become constrained superfields. Consistency of these
constraints with $\hat D^2 = D^2 \propto \partial_\tau$, implies that $K$ is a(n integrable)
complex structure on the space spanned by the twisted chiral fields, while $L$ should satisfy
one set of algebraic relations,
\begin{eqnarray}
K^k{}_l L^l{}_a = - L^k{}_b J^b{}_a, \label{int1}
\end{eqnarray}
and two sets of relations involving derivatives,
\begin{eqnarray}
0 &=& K^k{}_m K^m{}_{l,a} - K^k{}_m L^m{}_{a, l} + K^m{}_l L^k{}_{a, m} - L^m{}_a K^k{}_{l,m}
- J^b{}_a K^k{}_{l,b}, \nonumber \\
0 &=& K^k{}_l L^l{}_{[a,b]} + L^l{}_{[a} L^k{}_{b],l} + J^c{}_{[a} L^k{}_{b],c}.
\end{eqnarray}
This can be interpreted as follows. We can combine the constraints on the
chiral fields and eq.~(\ref{boc}) in a straightforward way to the
following constraint on the boundary,
\begin{eqnarray}
\hat D \left( \begin{array}{c}
            X^a \\
            X^k
            \end{array} \right)
            = \left( \begin{array}{cc}
            J^a{}_b & 0 \\
            L^k{}_b & K^k{}_l
            \end{array} \right)
            D\left( \begin{array}{c}
            X^b \\
            X^l
            \end{array} \right). \label{boc2}
\end{eqnarray}
The above conditions on $K$ and $L$ are then nothing but the statement that the matrix
\begin{eqnarray}
{\cal K} = \left( \begin{array}{cc}
            J & 0 \\
            L & K
            \end{array} \right) \label{calK}
\end{eqnarray}
represents a complex structure, ${\cal K}^2 = -1$ and ${\cal N}^M{}_{NK}({\cal K},{\cal K}) = 0$.
Here we introduced indices $M,\,N, \dots$ which run over both chiral Neumann directions and twisted
chiral directions. In terms of these, we can thus write
\begin{eqnarray}
\hat D X^M = {\cal K}^M{}_N DX^N. \label{boc3}
\end{eqnarray}

Because of (\ref{boc3}), the $X^M$ are not all independent. In order to deal with this when
considering the boundary term in variation of the action, these constraints are again solved
by introducing fermionic superfields $\tilde\Lambda^M$ such that
\begin{eqnarray}
\delta X^M = \frac{\partial X^M}{\partial \tilde X^N} \left( \hat D \delta\tilde\Lambda^N -
\tilde{\cal K}^N{}_P D\delta\tilde\Lambda^P \right), \label{consolv}
\end{eqnarray}
where $\tilde X$ are coordinates with respect to which $\tilde{\cal K}$ is
constant.\footnote{Note that the coordinates $\tilde X$ need not separate nicely into a
set of chiral and a set of twisted chiral superfields. In the end, the boundary term in the
variation will again be expressed in terms of the coordinates $X^{\tilde a}$ and $X^k$.} Note
that the chiral component of (\ref{consolv}) is nothing but
\begin{eqnarray}
\delta X^a = \hat D \delta \Lambda^a - J^a{}_b D \delta \Lambda^b, \label{consolv2}
\end{eqnarray}
where (because both $J^a{}_b$ and $\tilde J^a{}_b$ are constant)
\begin{eqnarray}
\delta\Lambda^a = \frac{\partial X^a}{\partial \tilde X^M} \delta\tilde\Lambda^M.
\end{eqnarray}
This shows that the $\delta\Lambda^a$ are exactly the unconstrained superfields needed to obtain
(\ref{bsfab}), so that this construction is consistent with previously obtained expressions. We
now write the second line of (\ref{bsfab}) as,
\begin{eqnarray}
 i \int d \tau \,d^2 \theta  \, M_k \delta X^k = i \int d \tau \,d^2 \theta  \, (M_N \delta X^N -
 M_a \delta X^a) \label{trick}
\end{eqnarray}
where we introduced $M_N = W_N + V_M J^M{}_N$. A calculation formally
identical to the one leading to eq.~(4.42) of \cite{Sevrin:2007yn} then
yields
\begin{eqnarray}
i \int d \tau \,d^2 \theta  \, M_N \delta X^N = i \int d \tau \,d^2 \theta  \, \delta \Lambda^M DX^N
\left( M_{M,P }{\cal K}^P{}_N - M_{P,N} {\cal K}^P{}_M + 2 M_P {\cal K}^P{}_{[N,M]} \right), \nonumber
\end{eqnarray}
where we used (\ref{consolv}) and we introduced the notation
\begin{eqnarray}
 \delta \Lambda^M =   \frac{\partial X^M}{\partial \tilde X^N} \delta \tilde\Lambda^N.
\end{eqnarray}
The second term of (\ref{trick}) is easier to work out in a similar way using (\ref{consolv2}).
Putting all pieces together, one arrives at the following expression for the boundary term in
the variation of the action (\ref{bsfab}):
\begin{eqnarray}
\delta {\cal S}\Big|_{boundary}&=& {\cal I}_A + {\cal I}_B,
\end{eqnarray}
where
\begin{eqnarray}
{\cal I}_A &=& -i \int d \tau \,d^2 \theta  \,
\delta \Lambda^a \left( \hat D' V_b  J^b{}_a + D' V_a - \hat DW_a + DW_b  J^b{}_a + DM_k  L^k{}_a
\right) \nonumber\\
&&-i \int d \tau \,d^2 \theta  \, \delta \Lambda^a \left( 2M_k L^k{}_{[a,b]} DX^b + M_l L^l{}_{a,k}
DX^k - M_l K^l{}_{k,a} DX^k  \right),
\end{eqnarray}
and
\begin{eqnarray}
{\cal I}_B &=& i \int d \tau \,d^2 \theta  \, \delta \Lambda^k \left( \hat D M_k - DM_l
K^l{}_k \right) +\nonumber\\
&&i \int d \tau \,d^2 \theta  \, \delta \Lambda^k \left( 2M_m K^m{}_{[l,k]} DX^l + M_l L^l{}_{a,k}
DX^a - M_l K^l{}_{k,a} DX^a  \right).
\end{eqnarray}
The first term ${\cal I}_A$ disappears when imposing Dirichlet conditions,
\begin{eqnarray}
\delta \Lambda^{\hat a} = 0, \label{piplusd}
\end{eqnarray}
and Neumann conditions,
\begin{eqnarray}
0 &=& \hat D' V_{\tilde b}  J^{\tilde b}{}_{\tilde a} + D' V_{\tilde a} - \hat DW_{\tilde a} +
DW_{\tilde b}  J^{\tilde b}{}_{\tilde a} + DM_k  L^k{}_{\tilde a}  \nonumber \\
&& +2M_k L^k{}_{[\tilde a,\tilde b]} DX^{\tilde b} + M_l ( L^l{}_{\tilde a,k} -
K^l{}_{k,\tilde a}) DX^k .\label{ntc1}
\end{eqnarray}
The second term ${\cal I}_B$ vanishes if we impose
 \begin{eqnarray}
0 &=& M_{k,m} K^m{}_l - M_{m,l} K^m{}_k + 2M_m K^m{}_{[l,k]}, \label{mc1}\\
0 &=& M_{k,{\tilde b}} J^{\tilde b}{}_{\tilde a} - M_{l,{\tilde a}} K^l{}_k + M_{k,l}
L^l{}_{\tilde a} + M_l L^l{}_{\tilde a,k} - M_l K^l{}_{k,\tilde a}. \label{mc2}
\end{eqnarray}
Using the fact that
\begin{eqnarray}
F_{kl} = -g_{km} (JK)^m{}_l, \label{Fkl}
\end{eqnarray}
and the boundary condition (\ref{mc1}), we find that $F_{kl} = \partial_k A_l - \partial_l A_k$, where
\begin{eqnarray}
A_k = \frac 1 2 M_l K^l{}_k + \partial_k f, \label{A_k}
\end{eqnarray}
with $f$ an arbitrary real function. This shows that the $U(1)$ gauge
fields in the twisted chiral directions are unaltered with respect to the
case where only twisted chiral fields are present \cite{Sevrin:2007yn}.
On the other hand, we have that
\begin{eqnarray}
F_{k{\tilde a}} = -g_{kl} (JL)^l{}_{\tilde a} - b_{k{\tilde a}}, \label{Fka}
\end{eqnarray}
where, again, the b-field is given by (\ref{KRform}). This together with (\ref{mc2}) implies
that $F_{k{\tilde a}} = \partial_k A_{\tilde a} - \partial_{\tilde a} A_k$, where $A_k$ is again
given by (\ref{A_k}) and $A_{\tilde a}$ can be written as,
\begin{eqnarray}
A_{\tilde a} = \frac 1 2 M_k L^k{}_{\tilde a} + \frac 1 2 M_{\tilde b} J^{\tilde b}{}_{\tilde a}
+ \partial_{\tilde a} f. \label{A_a}
\end{eqnarray}
Eqs.~(\ref{A_k}) and (\ref{A_a}) can be summarized by using the complex
structure $\cal K$,
\begin{eqnarray}
\left( \begin{array}{cc}
            A_{\tilde a} & A_k
            \end{array} \right)
            = - \frac 1 2 \left( \begin{array}{cc}
            M_{\tilde b} & M_l
            \end{array} \right)
            \left( \begin{array}{cc}
            J^{\tilde b}{}_{\tilde a} & 0 \\
            L^l{}_{\tilde a} & K^l{}_k
            \end{array} \right)
            + \left( \begin{array}{cc}
            \partial_{\tilde a} & \partial_k
            \end{array} \right) f\,,
\end{eqnarray}
or,
\begin{eqnarray}
A_M = \frac 1 2 M_N {\cal K}^N{}_M + \partial_M f. \label{A_M}
\end{eqnarray}
In terms of the field strength derived from (\ref{A_M}), the Neumann boundary conditions (\ref{ntc1})
can be rewritten as
\begin{eqnarray}
g_{\tilde a b} D' X^b = F_{\tilde a \tilde b} DX^{\tilde b} + (F_{\tilde a k} + b_{\tilde a k})
DX^k, \label{ntc2}
\end{eqnarray}
{\em i.e.} precisely of the general form (\ref{nbc9}). The non-standard boundary term (\ref{Eva1})
can on the other hand be rewritten as
\begin{eqnarray}
{\cal S}_{boundary}&=& i\,\int d \tau \,d \theta \, M_N \hat D X^N = 2i\,\int d \tau \,d \theta \, A_N DX^N,
\end{eqnarray}
where the last expression is obtained by using (\ref{boc3}) and (\ref{A_M}). This is precisely of
the standard form (\ref{Eva}).

Because of (\ref{Fkl}), both $F_{kl}$ and $\omega_{kl} = g_{km}
J^m{}_{l}$ -- which is anti-symmetric because the metric is hermitian
with respect to $J$, but is not closed when $H=db$ is non-trivial -- are
non-degenerate $(2,0) + (0,2)$ forms with respect to $K$. This implies
that the part of the target space spanned by the twisted chiral
superfields is $4l$-dimensional (with $l\in \mathbb{N}$ and $n=2l$ the
number of twisted chiral superfields). We conclude that
eqs.~(\ref{piplusn}), (\ref{piplusd}) and (\ref{ntc1}) describe a
$2(2l+k)$-dimensional brane on a $2(2l+m)$-dimensional target space. Note
that when no chiral fields are present -- $m=k=0$ -- we recover the
maximally coisotropic boundary conditions discussed in
\cite{Sevrin:2007yn}. We will therefore henceforth refer to this type of
boundary conditions as generalized coisotropic.

For a complete classification of D-branes on bihermitian geometries with two commuting complex structures,
it remains to discuss the more general setting where both $\pi_+$ and $\pi_-$ are nonzero. Note
however that -- since there will be $4l$ $\pi_+$-directions -- the lowest-dimensional example of such a brane
requires a six-dimensional target space, parameterized by twisted chiral fields exclusively. This case was
already considered in \cite{Sevrin:2007yn}. An example involving chiral fields as well, will necessarily require
a target space of eight dimensions or higher and will thus be physically less relevant. Because in a discussion
of the more general case the expressions would become far more complicated, we therefore only briefly outline
how more general boundary conditions can be obtained.

To this end, we assume the existence of adapted coordinates $X^{\check
\mu }$ and $X^{\hat \mu }$ (and their complex conjugates), $\check \mu ,
\check \nu , \cdots\in\{1,\cdots , l\}$ and $\hat \mu , \hat \nu ,
\cdots\in\{l+1,\cdots , n\}$, such that the only non-vanishing components
of $ \pi _+$ and $ \pi _-$ are $ \pi _-^{\hat \mu }{}_{ \hat \nu }=
\delta ^{ \hat \mu }_{ \hat \nu }$ and $ \pi_+^{\check \mu }{}_{\check
\nu }= \delta ^{ \check \mu }_{\check \nu }$. With this,
eqs.~(\ref{dcl3}) and (\ref{dcl4}) become,
\begin{eqnarray}
\delta X^ { \hat \mu } &=& R^ { \hat \mu} {}_{ \bar { \hat\nu} }\,
\delta X^{ \bar{ \hat \nu} }+ R^ { \hat\mu} {}_ {\tilde \beta }\,
\delta X^{ \tilde \beta }+
R^ { \hat\mu} {}_{\bar{\tilde \beta }}\, \delta X^{\bar{\tilde \beta }}, \label{dcl6}\\
D'X^ { \hat\mu} &=&
-R^{ \hat \mu} {}_{ \bar { \hat\nu} }\, D' X^{ \bar{ \hat \nu} }+
R^ { \hat\mu} {}_ {\tilde \beta }\, D X^{ \tilde \beta }-
R^ { \hat\mu} {}_{\bar{\tilde \beta }}\, D X^{\bar{\tilde \beta }},\label{dcl5}
\end{eqnarray}
while (\ref{piplusn}) becomes
\begin{eqnarray}
D'X^{\check k} = F^{\check k}{}_l DX^l + (F^{\check k}{}_{\tilde a} + b^{\check k}{}_{\tilde a})
D X^{\tilde a}. \label{dcl7}
\end{eqnarray}
Implementation of this in eq.~(\ref{bsfab}) then yields the appropriate
boundary conditions on the chiral fields. These will either be of the
standard Dirichlet form (\ref{dcl1}) or of a form which will ultimately
be equivalent to the Neumann conditions (\ref{nbc9}). One way to find the
exact form of these Neumann conditions, is to go to worldvolume
coordinates on the brane. These coordinates should be such that
eq.~(\ref{dcl6}) is trivially satisfied. The field strengths which should
enter in these Neumann conditions, can be obtained by using (\ref{dcl5})
and (\ref{dcl7}) in (\ref{Eva1}) and comparing the result to the standard
form of the $N=1$ boundary term (\ref{Eva}).

In this way a wide variety of branes can be obtained. For example, for an
8-dimensional target space described by one chiral field and three
twisted chiral fields, one obtains D3- and D5-branes if $\pi_- = 1$, and
D5- and D7-branes when $\pi_- \neq 1$ (note that in this case $\pi_+$
cannot equal 1, because this can only happen when there are an even
number of twisted chiral fields). The D5- and D7-branes occur when
generalized coisotropic Neumann conditions are imposed on two of the
twisted chiral fields.

\section{Examples} \label{section:ex}
\subsection{A four-dimensional target manifold}
\subsubsection{Generalities}
We consider a 4-dimensional target manifold. We can distinguish four different cases.
\begin{itemize}
\item It is parameterized in terms of two chiral superfields.\\
This case was studied in \cite{Sevrin:2007yn}. It describes either a
D0-, or a D2- or a D4-brane on a K{\"a}hler manifold wrapping around a
holomorphic cycle. It also goes under the name of a B-brane on a
K{\"a}hler manifold.
\item It is parameterized in terms of two twisted chiral superfields.\\
This case was also studied in \cite{Sevrin:2007yn}. It
describes either a D2-brane wrapped on a lagrangian submanifold of a
K{\"a}hler manifold or a maximally coisotropic D4-brane on a K{\"a}hler
manifold. These branes also go under the name of A-branes.
\item It is parameterized in terms one chiral and one twisted chiral superfield.\\
This is the case we will study next. As we will show we can either describe D1- or D3-branes on a
bihermitian manifold with commuting complex structures.
\item It is parameterized in terms of a semi-chiral multiplet.\\
This case will be studied elsewhere \cite{wip}. However we will
briefly touch upon it when discussing duality transformations in
section 6. Here it is sufficient to say that this case corresponds to
either a D2- or a D4-brane on a bihermitian geometry where $\ker
[J_+,J_-]=\emptyset$.
\end{itemize}

Let us now focus on the case where one has one chiral superfield $z$ and
one twisted chiral superfield $ w$. The boundary term eq.~(\ref{bsfab})
becomes,
\begin{eqnarray}
\delta {\cal S}\Big|_{boundary}&=&\int d \tau \,d^2 \theta  \,
\Big\{
\delta \Lambda  \big( \bar \ID'V_ { z }+i\, \bar \ID W_{ z }
\big)-
\delta \bar \Lambda  \big(
 \ID'V_ { \bar z }-i\,  \ID W_{ \bar z }\big)- \nonumber\\
&&\qquad
 \delta w \big(V_{ w }-i\, W_{ w }\big)+
 \delta \bar w  \big(V_{ \bar w }+i\, W_{ \bar w }\big)
\Big\}.\label{bsfabex}
\end{eqnarray}
By choosing appropriate boundary conditions this term should vanish. We
have only a single chiral field which leads us immediately to two
subcases: either we impose Dirichlet boundary conditions in the chiral
direction or not. For the twisted chiral superfield we will always have a
Dirichlet and a Neumann boundary condition. So having Dirichlet boundary
conditions in the chiral direction will lead to a D1-brane while Neumann
boundary conditions in the chiral direction gives a D3-brane. Instead of
dwelling on the general case we will focus on a few concrete examples
which highlight all subtleties.

\subsubsection{D3-branes on $T^4$}\label{subsubsection:t4}
We parametrize $T^4$ by a chiral $z$ and a twisted chiral $w$ coordinate.
For simplicity we choose the torus such that,
\begin{eqnarray}
 z\simeq z+\frac{1}{\sqrt{2}}\big(\IZ+i\,\IZ\big),\qquad
  w\simeq w+\frac{1}{\sqrt{2}}\big(\IZ+i\,\IZ\big).\label{t41}
\end{eqnarray}
Note that one easily generalizes the present analysis to an arbitrary
point in the moduli space of $T^4$. The generalized K{\"a}hler potential is
simply given by,
\begin{eqnarray}
 V=z\bar z-w\bar w.\label{van1}
\end{eqnarray}
As mentioned here above we can have either D1- or D3-branes. The former
is not particularly interesting as by setting $z$ to a constant we end up
with a D1-brane wrapping around a lagrangian submanifold of the 2-torus
(a trivial concept in two dimensions) parametrized by $w$ which was
already discussed in \cite{Sevrin:2007yn}. So we turn to the D3-brane and
we choose a simple linear Dirichlet boundary condition,
\begin{eqnarray}
 \alpha \,w+\bar \alpha \,\bar w=\beta \,z+\bar \beta \,\bar z,\label{t42}
\end{eqnarray}
where $ \alpha,\,\beta \in\IC$ and $\alpha \neq 0$. Consistency of this
with eq.~(\ref{t41}) requires that $\alpha ,\,\beta \in\IZ+i\,\IZ$. In the language of
subsection \ref{subs:bc}, this corresponds to taking
\begin{eqnarray}
R^z{}_z=1,\quad R^w{}_{\bar w} = - \frac{\bar\alpha}{\alpha},\quad R^w{}_{z} =
\frac{\beta}{\alpha}, \quad R^w{}_{\bar z} = \frac{\bar\beta}{\alpha}. \label{RT4}
\end{eqnarray}
Because of eqs.~(\ref{bcf}) and (\ref{btcf}), eq.~(\ref{t42}) immediately
implies the Neumann boundary condition,
\begin{eqnarray}
 \alpha \,D'w-\bar \alpha \,D'\bar w=\beta \,Dz-\bar \beta \,D\bar z.\label{t43}
\end{eqnarray}
Requiring now that the terms proportional to $\delta w$ and $\delta \bar
w$ in the boundary term in the variation of the action,
eq.~(\ref{bsfabex}), vanish yields the boundary potential,
\begin{eqnarray}
 W= \frac{i}{2}\frac{\alpha }{\bar \alpha }\,w^2-\frac{i}{2}\frac{\bar \alpha }{\alpha }\,\bar w^2+f(z,\bar z),\label{t45}
\end{eqnarray}
with $f(z,\bar z)$ an arbitrary\footnote{It is clear that this function
should obey appropriate periodicity conditions consistent with the global
properties of the torus, {\em i.e.} $f\big(z+(m+i\,n)/\sqrt{2},\bar
z+(m-i\,n)/\sqrt{2}\big)=f(z,\bar z)+h(z)+\bar h(\bar z)$ with $h(z)$ an
arbitrary holomorphic function and $m,n\in\IZ $.} real function of $z$
and $\bar z$, and two more Neumann boundary conditions,
\begin{eqnarray}
 \ID'z&=&- \frac{\bar \beta }{\bar \alpha }\,\ID w+\frac{\bar \beta }{\alpha }\,\ID\bar w+i\,f_{z\bar z}\,\ID z,\nonumber\\
 \bar \ID'\bar z&=&+ \frac{\beta }{\bar \alpha }\,\bar \ID w-\frac{ \beta }{\alpha }\,\bar \ID\bar w-
 i\,f_{z\bar z}\,\bar \ID \bar z.
 \label{t44}
\end{eqnarray}
This is indeed of the general form (\ref{piminn2}) for an almost product
structure given by (\ref{RT4}). So we end up with a D3-brane whose
position is given by eq.~(\ref{t42}). Using eq.~(\ref{t43}) in
eq.~(\ref{Eva1}) and comparing the result to eq.~(\ref{Eva}) we identify
the $U(1)$ gauge fields,
\begin{eqnarray}
&& A_z=\frac 1 2 \left(i\,f_z- \frac{\beta }{\bar \alpha }\,w+\frac{\beta }{\alpha }\,\bar w\right),\nonumber\\
&& A_{\bar z}=\frac 1 2 \left(-i\,f_{\bar z}+ \frac{\bar \beta }{\bar \alpha }\,w-
\frac{\bar \beta }{\alpha }\,\bar w\right),\nonumber\\
&& A_w=A_{\bar w}=0,
\end{eqnarray}
as anticipated in eqs.~(\ref{piminA}).

As mentioned in subsection \ref{subs:bc}, one can show that when using worldvolume coordinates,
the Neumann boundary conditions (\ref{t43}) and (\ref{t44}) reduce to a more familiar form.
Writing $\alpha = m_1 + i\, m_2$ and $\beta = m_3 + i\, m_4$, where
$m_i\, \in \, \IZ$ and assuming $m_2 \neq 0$, we introduce the
worldvolume coordinates $r$, $s$ and $t$, and write,
\begin{eqnarray}
z=r+i\,s, \quad w = \left(1+i\, \frac{m_1}{m_2} \right) t - i\,
\frac{m_3}{m_2}\, r + i\, \frac{m_4}{m_2}\, s.
\end{eqnarray}
The non-vanishing components of the pullback of the $U(1)$ fieldstrength
are then given by,
\begin{eqnarray}
F_{rs} = -2\,\frac{m_1}{m_2}\,\frac{m_3^2 + m_4^2}{m_1^2 + m_2^2} -
\frac 1 2(f_{rr} + f_{ss})\,, \quad F_{rt} = -2 \,\frac{m_4}{m_2}\,, \quad F_{st}= -2\,
\frac{m_3}{m_2}\,.
\end{eqnarray}
Since there is no torsion present in this example, the U(1) field
strength $F$ and the invariant two-form ${\cal F} = F + b$ are equal to
each other. Using the above expressions for the components of $F$, it is
not hard to show that eqs.~(\ref{t43}) and (\ref{t44}) are equivalent to
the standard Neumann boundary conditions (\ref{nbc9}).

\subsubsection{$S^3\times S^1$}\label{subsection:S3S1} Wess-Zumino-Witten
models are non-trivial but still relatively simple examples of non-linear
$ \sigma $-models with $N\geq (2,2)$ \cite{Spindel:1988nh}. The simplest
case is the WZW-model on $SU(2)\times U(1)$ (or the Hopf surface
$S^3\times S^1$) which is the only $N=(2,2)$ WZW-model which can be
parameterized without the use of semi-chiral superfields. Parameterizing
the WZW group element as,
\begin{eqnarray}
{\cal G}= \frac{e^{-i\, \ln \sqrt{z \bar z+ w \bar w} }}{\sqrt{z
\bar z+ w \bar w}}\,\left(
\begin{array}{cc} w & \bar z\\
-z & \bar w\end{array}\right),
\end{eqnarray}
it was found in \cite{Rocek:1991vk}, \cite{Rocek:1991az},
that the generalized K{\"a}hler potential is
explicitly given by,
\begin{eqnarray}
V&=&-\int^{w \bar w/z \bar z} \frac{dq}{q}\ln\big(1+q\big)+
\frac 1 2 \big(\ln z\, \bar z\big)^2 \nonumber\\
&=&+\int^{z \bar z/w \bar w} \frac{dq}{q}\ln\big(1+q\big)-
\frac 1 2 \big(\ln w\, \bar w\big)^2+\ln(w \bar w)\ln(z \bar z).\label{WZWpot1}
\end{eqnarray}
Note that the equality $V(z, \bar z,w, \bar w)=-V(w, \bar w, z, \bar z)$
holds modulo a generalized K{\"a}hler transformation. The potential
eq.~(\ref{WZWpot1}) correctly encodes the metric and the torsion of the
group manifold (see eq.~(\ref{KRform})). Parameterizing $SU(2)\times
U(1)$ with Hopf coordinates $z=\cos \psi \,e^{ \rho +i \phi _1}$, $w=\sin
\psi \,e^{ \rho + i\phi _2}$, with $ \phi _1,\, \phi_2,\, \rho \in
\IR\,\mbox{mod}\,2 \pi  $ and $\psi\in[0, \pi /2]$, we find that $z,
\,w\in (\IC^2\setminus 0)/\Gamma$ where $\Gamma$ is generated by
$(z,w)\rightarrow (e^{2 \pi }\,z,e^{2 \pi }\,w)$ which is precisely the
definition of a Hopf surface.

We will use the bulk potential,
\begin{eqnarray}
V=+\int^{z \bar z/w \bar w} \frac{dq}{q}\ln\big(1+q\big)-
\frac 1 2 \big(\ln w\, \bar w\big)^2,\label{WZWpot2}
\end{eqnarray}
which differs from eq.~(\ref{WZWpot1}) by a generalized K{\"a}hler transformation.
In addition we have that global consistency requires invariance under,
\begin{eqnarray}
z\rightarrow e^{2 \pi n}\,z,\quad w\rightarrow e^{2 \pi n}\,w,\qquad n\in\IZ.
\end{eqnarray}
Under this the generalized K{\"a}hler potential transforms as,
\begin{eqnarray}
V\rightarrow V -4 \pi\, n \ln( w \bar w)-8 \pi ^2n^2,
\end{eqnarray}
which is a generalized K{\"a}hler transformation eq.~(\ref{genKahltrsf1}).
In order to restore the invariance the boundary potential should transform as well (see
eq.~(\ref{genKahlbd})),
\begin{eqnarray}
W\rightarrow W+4 \pi\, n\, i\ln\left( \frac{w}{ \bar w}\right),\label{Eva3}
\end{eqnarray}
which should hold modulo the sum of a holomorphic and an anti-holomorphic
function of the chiral fields (see eqs.~(\ref{awq1}-\ref{awq2})).

\vspace{.3cm}

\noindent\underline{i. D1-branes}\\
We first study D1-branes. We
impose the Dirichlet boundary conditions,
\begin{eqnarray}
&&z=z_0,\qquad \bar z= \bar z_0, \nonumber\\
&&-i \ln \frac{w}{ \bar w}=Q'\big(\ln(z_0 \bar z_0+ w \bar w)\big),\label{Eva4}
\end{eqnarray}
where we parameterized the boundary potential $W\big(\ln(z_0 \bar z_0+ w \bar w)\big)$ as,
\begin{eqnarray}
W(x)=Q(x)-x\,Q'(x),
\end{eqnarray}
where $Q'(x)\equiv \partial _x Q(x)$ and $x\equiv\ln(z_0 \bar z_0+ w \bar
w)$. Requiring this to be consistent with eq.~(\ref{Eva3}) gives,
\begin{eqnarray}
Q(x)=f\big(\sin( \frac{x}{2})\big)+ \frac{m}{2}\,x^2+ a\,x,\label{QD1}
\end{eqnarray}
with $f(y)\in\IR$ an arbitrary function and $m,\,a\in\IR$. Furthermore
requiring that the periodicity of the left hand side of the last equation
in (\ref{Eva4}) is correctly reproduced by the right hand side forces us
to take $m\in\IZ$. One recognizes the integer $m$ as the winding number
in the $\phi _2$ direction, {\em i.e.} going once around the circle
parameterized by $ \rho $ one winds $m$ times around the circle
parameterized by $\phi _2$.

\vspace{.3cm}

\noindent\underline{ii. D3-branes}\\
We now turn to the D3-brane. We
introduce some notation,
\begin{eqnarray}
 x\equiv \ln ( z\bar z+w\bar w),\qquad
 y\equiv -i\,\ln\frac{ z}{ \bar z},
\end{eqnarray}
and we denote a derivative w.r.t. $x$ by a prime. Parameterizing the
boundary potential $W(x,z, \bar z)$ as,
\begin{eqnarray}
 W(x,z, \bar z)=Q(x,z, \bar z)-x\,Q'(x,z, \bar z)
\end{eqnarray}
and imposing a single Dirichlet boundary condition,
\begin{eqnarray}
-i \ln \frac{w}{ \bar w}=Q'(x,z, \bar z),\label{EV1}
\end{eqnarray}
we get that the terms in eq.~(\ref{bsfabex}) which are proportional to
$\delta w$ and $\delta \bar w$ cancel. Using eq.~(\ref{EV1}) in
(\ref{Eva3}) we get that,
\begin{eqnarray}
 W\big(x+4\pi n,e^{2\pi n}z,e^{2\pi n}\bar z\big)=W(x,z,\bar z)-4\pi n\,Q'(x,z,\bar z),\label{eto1}
\end{eqnarray}
should hold. Taking once more a derivative of this with respect to $x$,
we find that $Q''(x,z,\bar z)$ is periodic,
\begin{eqnarray}
 Q''\big(x+4\pi n,e^{2\pi n}z,e^{2\pi n}\bar z\big)=Q''(x,z,\bar z).
\end{eqnarray}
In principle we could solve this and integrate it to $Q$ while
implementing eq.~(\ref{eto1}). However we will limit ourselves here to a
simple choice which satisfies all requirements,
\begin{eqnarray}
 Q=\frac{m_1}{2}\,x^2+m_2\,y\,\big(x-\ln z\bar z\big).
\end{eqnarray}
 With this the boundary potential becomes,
\begin{eqnarray}
 W=-\frac{m_1}{2}\,x^2-m_2\,y\,\ln z\bar z,
\end{eqnarray}
and the Dirichlet boundary condition is explicitly,
\begin{eqnarray}
-i \ln \frac{w}{ \bar w}=m_1\,x+m_2\,y.\label{EV2}
\end{eqnarray}
As $-i\ln w/\bar w$, $x$ and $y$ are all periodic we get that
$m_1,\,m_2\in\IZ$. In the language of subsection \ref{subs:bc}, this
corresponds to $R^z{}_z=1$ and,
\begin{eqnarray}
R^w{}_{\bar w} &=& \frac{w}{\bar w}\; \frac{z \bar z + (1+im_1) w \bar w}{z \bar z + (1-im_1) w \bar w},\nonumber\\
R^w{}_{z} &=& \frac{w}{z}\; \frac{(m_2+im_1) z \bar z + m_2 w \bar w}{z \bar z + (1-im_1) w \bar w},  \label{Rhopf}\\
R^w{}_{\bar z} &=& -\frac{w}{\bar z}\; \frac{(m_2-im_1) z \bar z + m_2 w \bar w}{z \bar z + (1-im_1) w \bar w}.\nonumber
\end{eqnarray}
The Dirichlet boundary condition implies a Neumann
boundary condition as well,
\begin{eqnarray}
 \ID'\ln w\bar w=i\,m_1\,\ID x+i\,m_2\,\ID y.\label{EV69}
\end{eqnarray}
Using eq.~(\ref{EV2}) in the boundary term in the variation of the
action, eq.~(\ref{bsfabex}), one finds that the remaining terms
proportional to $\delta \Lambda $ and $\delta \bar \Lambda $ vanish
provided two more Neumann boundary conditions are imposed,
\begin{eqnarray}
 \ID'z&=&i\,m_1\,z\,\ID x-\frac{m_2}{\bar z}\,\ID\big(w\bar w\big),\nonumber\\
\bar \ID'\bar z&=&-i\,m_1\,\bar z\,\bar \ID x-\frac{m_2}{z}\,\bar \ID\big(w\bar w\big).\label{EV3}
\end{eqnarray}
These are indeed equivalent to (\ref{piminn2}) for an almost product structure given by (\ref{Rhopf}),
once the other Neumann condition (\ref{EV69}) is imposed.
Using eq.~(\ref{EV69}) in eq.~(\ref{Eva1}) and comparing it to
eq.~(\ref{Eva}) leads to a $U(1)$ bundle with potential fields,
\begin{eqnarray}
A_w &=& A_{\bar w} = 0, \nonumber \\
A_z &=& -\frac{1}{2}\left( V_z-m_2\,\frac{x}{z}\right), \qquad A_{\bar z} = - \frac{1}{2}
\left(V_{\bar z}-m_2\,\frac{x}{\bar z}\right),
\end{eqnarray}
again in agreement with eq.~(\ref{piminA}). Introducing world volume
coordinates $\rho$, $\psi$ and $\phi$ such that,
\begin{eqnarray}
 z=\cos\psi\,e^{\rho+i\,\phi},\qquad
 w=\sin\psi\,e^{(1+i\,m_1)\rho+i\,m_2\,\phi},
\end{eqnarray}
we find that the only non-trivial component of the U(1) bundle pullback
to the D3-brane is given by $F_{\rho \psi} = -2
\big(\cot{\psi}+m_2\,\tan{\psi}\big)$. Combining with the NS-NS 2-form
pullback to the D3-brane we obtain the pullback of the invariant 2-form
${\cal F} = b + F$,
\begin{eqnarray}
{\cal F}_{\rho \phi } = -2\, m_1\, \cos^2 \psi,\qquad
{\cal F}_{\rho \psi } = -2\, m_2\, \tan \psi.
\end{eqnarray}
Given this (and the pullback of the metric to the worldvolume), we expect Neumann boundary conditions of the
form (see (\ref{nbc9})),
\begin{eqnarray}
&(1+m_1^2 \sin^2 \psi)\, D'\rho + m_1 m_2 \sin^2\psi\, D'\phi = - m_2 \tan \psi \, D\psi -
m_1 \cos^2 \psi \, D\phi,& \nonumber \\
&(1 + m_2^2 \tan^2 \psi)\, D'\phi + m_1 m_2 \tan^2 \psi \, D'\rho = m_1 D\rho,& \\
&D' \psi = m_2 \tan \psi\, D\rho,& \nonumber
\end{eqnarray}
which is indeed equivalent to eqs.~(\ref{EV69}) and (\ref{EV3}). It would
be an instructive exercise to repeat the analysis of \cite{Bachas:2000ik}
for this particular manifest $N=2$ supersymmetric case.

\subsection{New space-filling branes on $T^6$}\label{subsection:D6T6}
To illustrate the case $\pi_+ = 1$ of subsection (\ref{subs:bc}), we now
discuss the (purely Neumann) boundary conditions for a space filling
brane on $T^6$ parameterized by one chiral superfield $z$ and two twisted
chiral superfields $w^\mu$, $\mu \in \{1,2\}$, where we impose
generalized coisotropic boundary conditions on the twisted chiral fields.
Consider the potential
\begin{eqnarray}
V = z \bar z - w^1\bar w^1 - w^2 \bar w^2 + b_1 (z\bar w^1 + \bar z w^1) + b_2 (z\bar w^2 + \bar z w^2), \label{VD6T6}
\end{eqnarray}
with constant $b_\mu \in \mathbb{R}$, $\mu \in \{1,2\}$, so that according to (\ref{KRform}) the
b-field has nonzero components $b_{\mu\bar z} = b_\mu$. By virtue of (\ref{Fkl}) and (\ref{Fka})
this implies the following relation between the $U(1)$ field strength and the components of the
complex structure (\ref{calK})
\begin{eqnarray}
F_{\mu\nu} = i K^{\bar\mu}{}_\nu, &&\quad F_{\mu\bar\nu} = i K^{\bar\mu}{}_{\bar\nu},\nonumber\\
F_{\mu z} = i L^{\bar\mu}{}_z, &&\quad F_{\mu\bar z} = i L^{\bar\mu}{}_{\bar z} - b_\mu. \label{FvsKL}
\end{eqnarray}
{From} (\ref{Fkl}) it follows that the complex structure $K^k{}_l$ cannot
be proportional to $J^k{}_l$. A good choice is $K^1{}_{\bar 2} = 1$,
$K^2{}_{\bar 1} = -1$, and other unrelated components zero. This
corresponds to the choice
\begin{eqnarray}
F_{12} = i,  \quad F_{1\bar 2} = 0. \label{Ftt}
\end{eqnarray}
On the other hand, the field strength can be computed from the $U(1)$
potentials (\ref{A_M}). Assuming a quadratic form of $W$ (so that its
second derivatives $W_{MN}$ are constants) and with the above choice of
$K$, this implies a relation between components of $L$ and second
derivatives of $W$. It turns out that we can find a
non-trivial\footnote{Note that putting $L$ and $W$ to zero results in a
rather trivial example in the sense that it corresponds to a
4-dimensional maximally coisotropic system of the kind studied in
\cite{Sevrin:2007yn} along with two chiral spectator directions trivially
wrapped by the brane.} solution if $W_{\mu\nu} = W_{\mu\bar\nu} = 0$. In
that case, we find that we have to satisfy the following relations
\begin{eqnarray}
L^{\bar 1}{}_z &=& W_{1z} - b_2 -i W_{\bar 2 z},\nonumber \\
L^{\bar 2}{}_z &=& W_{2z} + b_1 +i W_{\bar 1 z}.  \label{LvsW}
\end{eqnarray}
while the other components of $L$ are fixed by eq.~(\ref{int1}) to be
\begin{eqnarray}
L^{1}{}_z = i L^{\bar 2}{}_z, \quad L^{2}{}_z = -i L^{\bar 1}{}_z,
\end{eqnarray}
so that $L$ is fully determined by specifying {\em e.g.} $L^{\bar 1}{}_z$
and $L^{\bar 2}{}_z$. Let us take $W_{1z}$, $W_{2z}$, $W_{\bar 1z}$,
$W_{\bar 2z}\in\IR$ and write the left hand side of (\ref{LvsW}) as
\begin{eqnarray}
L^{\bar 1}{}_z &\equiv& \alpha = \alpha_1 + i \alpha_2, \nonumber \\
L^{\bar 2}{}_z &\equiv& \beta = \beta_2 + i\beta_1,
\end{eqnarray}
with $\alpha _j$ and $\beta _j$, $j\in\{1,2\}$ real so that (\ref{LvsW})
is solved by
\begin{eqnarray}
\alpha_1 = W_{1z} - b_2, && \alpha_2 = - W_{\bar 2 z}, \nonumber \\
\beta_2 = W_{2z} + b_1, && \beta_1 = W_{\bar 1 z},
\end{eqnarray}
or
\begin{eqnarray}
W &=& (\alpha_1 + b_2)(w^1 z + \bar w^1 \bar z) + \beta_1 (w^1 \bar z + \bar w^1 z) + \nonumber \\
&& (\beta_2 - b_1) (w^2 z + \bar w^2 \bar z) -\alpha_2 (w^2 \bar z + \bar w^2 z) + f(z,\bar z), \label{WD6T62}
\end{eqnarray}
where $f(z,\bar z)$ is a real function. With this choice for $L$ and $W$, the components of the $U(1)$
gauge field become (up to gauge transformations)
\begin{eqnarray}
A_1 &=& -\frac 1 2 (\beta_2 - b_1) \bar z + \frac 1 2 (\alpha_2 + ib_2) z - \frac i 2 w^2, \nonumber \\
A_2 &=& +\frac 1 2 (\alpha_1 + b_2) \bar z + \frac 1 2 (\beta_1 - ib_1) z + \frac i 2 w^1 ,
\end{eqnarray}
and
\begin{eqnarray}
A_z &=& \frac 1 2 [\beta (\beta - 2b_1) + \alpha (\alpha + 2b_2)] \bar z + \frac i 2 f_z \nonumber \\
&+& \frac 1 2 (2i\alpha_1 - \alpha_2 + ib_2) w^1 + \frac 1 2 (2i\beta_2 - \beta_1 -ib_1) w^2 \\
&+& \frac 1 2 (\beta_2 +2i \beta_1 - b_1) \bar w^1 - \frac 1 2 (\alpha_1 +2i \alpha_2 + b_2) \bar w^2 ,\nonumber
\end{eqnarray}
and their complex conjugates. These indeed yield the required components
of the invariant field strength (see eq.~(\ref{FvsKL}))
\begin{eqnarray}
F_{1z} = i\alpha, && F_{1 \bar z} + b_{1 \bar z}  = \bar\beta,  \nonumber\\
F_{2z} = i\beta, && F_{2 \bar z} + b_{2 \bar z} = -\bar\alpha,
\end{eqnarray}
while for $F_{z\bar z}$ we find
\begin{eqnarray}
F_{z \bar z} = -i \alpha_2 (\alpha_1 + 2b_2) -i\beta_1(\beta_2 - 2b_1) - i W_{z\bar z},
\end{eqnarray}
which shows that for certain solutions, $F_{z\bar z}$ will depend not only on $f(z, \bar z)$, but
also on $L$ and the b-field. Note that the choice $\alpha_1 = -b_2$, $\beta_2 = b_1$ and
$\alpha_2 = \beta_1 = 0$, leads to a solution where the boundary potential $W$ is trivial
(modulo a term $f(z,\bar z)$), while $L$ is nontrivial and $F_{1z} = -ib_2$ and $F_{2z} = ib_1$.
This situation should be contrasted with the case where the b-field vanishes. In that case $W$
necessarily has to be nontrivial for $L$ to be nontrivial (as can be seen from (\ref{LvsW})).

An explicit construction of this kind of space-filling D6-brane on a more
non-trivial target space -- the simplest candidate being $S^3 \times S^1
\times T^2$ -- should in principle be possible. Furthermore, the solution
of this subsection should be dual to a coisotropic D5-brane on $T^6$ and
it should be possible to make this duality explicit by the methods
developed in the following section. We leave these matters for further
investigation.

\section{Duality transformations}
\subsection{Generalities}
T-duality transformations in $N=(2,2)$ supersymmetric non-linear $ \sigma
$-models correspond to duality transformations which interchange the
different types of superfields \cite{Gates:1984nk}, \cite{Rocek:1991vk},
\cite{Grisaru:1997ep}, \cite{Bogaerts:1999jc}, \cite{Lindstrom:2007sq},
\cite{Merrell:2007sr}. The simplest ones are those that allow to exchange
a chiral for a twisted chiral superfield and vice-versa when an isometry
is present. Gauging the isometry, one imposes -- using Lagrange
multipliers -- that the gauge fields are pure gauge. In this way,
integrating over the Lagrange multipliers gives back the original model.
However when integrating over the gauge fields (or their potentials which
are unconstrained superfields) one obtains the dual model.

Let us briefly review the case without boundaries. As a starting point we take the
action,
\begin{eqnarray}
{\cal S}_{(1)}=4\int d^2 \sigma \,d^4 \theta \left(
-\int^Y dq\, W(q,\cdots)+(z+ \bar z)\,Y
\right),\label{fstor}
\end{eqnarray}
where $Y$ is an unconstrained $N=(2,2)$ superfield, $z$ is either a
chiral or a twisted chiral superfield and $\cdots$ stands for other,
spectator fields. The equations of motion for $Y$ give,
\begin{eqnarray}
z+ \bar z=W(Y,\cdots ),
\end{eqnarray}
which upon inversion gives,
\begin{eqnarray}
Y=U( z+ \bar z,\cdots).
\end{eqnarray}
Using this to eliminate $Y$ yields the second order dual action,
\begin{eqnarray}
{\cal S}_{dual}= 4\int d^2 \sigma\, d^4 \theta\,\int^{z+ \bar z} dq\,U(q,\cdots).
\end{eqnarray}
Take now $z$ and $ \bar z$ to be chiral superfields and varying them yields,
\begin{eqnarray}
\bar \ID_+ \bar \ID_-Y=\ID_+\ID_-Y=0,
\end{eqnarray}
which is solved by putting $Y= w+ \bar w$ with $w$ a twisted chiral
superfield. If on the other hand we started off with a field $z$ which
was twisted chiral we get upon integrating over $z$ and $ \bar z$,
\begin{eqnarray}
\bar \ID_+  \ID_-Y=\ID_+\bar\ID_-Y=0,
\end{eqnarray}
which is now solved by putting $Y = w + \bar w$, with $w$ a chiral
superfield. The resulting second order action (which is the action one
starts with) is in both cases given by,
\begin{eqnarray}
{\cal S}=-4\int d^2 \sigma \,d^4 \theta \,\int^{w+ \bar w} dq\, W(q,\cdots).
\end{eqnarray}

Let us illustrate this using the previous example, the WZW model on the
Hopf surface $S^3\times S^1$. We will first dualize the twisted chiral
field to a chiral one. In order to do this we rewrite the potential
eq.~(\ref{WZWpot2}) as,
\begin{eqnarray}
 V=-\int^{w\bar w}\, \frac{dq}{q}\,\ln\big(q+z \bar z\big).
\end{eqnarray}
With this the first order action is given by,
\begin{eqnarray}
{\cal S}_{(1)}=-4\,\int\,d^2 \sigma \,d^2\theta \,d^2 \hat \theta \,
 \left\{\int^{e^Y}\frac{dq}{q}\ln\big(q+z\bar z\big)+Y\,\ln z' \bar z'\right\},
\end{eqnarray}
where $Y$ is an unconstrained superfield and $z'$ is a chiral superfield.
Integrating over $\ln z'$ and $ \ln \bar z'$ gives the original model
back. Integrating over $Y$ gives the dual model with action,
\begin{eqnarray}
 {\cal S}_{dual}=-4\,\int\,d^2 \sigma \,d^2\theta \,d^2 \hat \theta \,
 \left\{\int^{z''\bar z''}\frac{dq}{q}\ln\big(1-q\big)-\frac 1 2 \left(\ln z' \bar z'\right)^2\right\},
\end{eqnarray}
where we performed a change of coordinates $z\rightarrow z''=z\,z'$ .
Following the duality transformation in detail we find that the chiral
fields satisfy $|z''|\leq 1$ and $z'\simeq e^{2\pi (n_1+ i n_2)}\,z'$
with $n_1,\,n_2\in\IZ$. So the target manifold of the dual model
factorizes as a product of a disk (with a singular metric) and a torus.
\footnote{Note however that a proper treatment of this duality
transformation requires the presence of a non-trivial dilaton field in
the dual model \cite{Rocek:1991vk}. Indeed, if this were not the case we
would have expected that the dual model is hyper-K{\"a}hler which it is not.}

We now dualize the chiral field to a twisted chiral field. The first
order action is given by,
\begin{eqnarray}
{\cal S}_{(1)}=4\int\,d^2 \sigma \,d^2\theta \,d^2 \hat \theta \,
 \left\{\int^{e^Y}\frac{dq}{q}\ln\left(1+
\frac{q}{w\bar w} \right)-\frac 1 2 \left(\ln w \bar w\right)^2-Y\,\ln w' \bar w'\right\},
\end{eqnarray}
where $w'$ is twisted chiral and $Y$ is an unconstrained superfield. The
original model is recovered by integrating over $\ln w'$ and $\ln \bar
w'$. Integrating over $Y$ gives the dual model,
\begin{eqnarray}
{\cal S}_{dual}=4\int\,d^2 \sigma \,d^2\theta \,d^2 \hat \theta \,
 \left\{\int^{w''\bar w''}\frac{dq}{q}\ln\big(1-q\big)-\frac 1 2 \left(\ln w''' \bar w'''\right)^2\right\},
\end{eqnarray}
where we performed the following coordinate transformations,
\begin{eqnarray}
 w''= \frac{1}{w'}\,,\qquad w'''=w\,w'.
\end{eqnarray}
One finds that $|w''|\leq 1$ and $w'''\simeq e^{2\pi (n_1+ i n_2)}w'''$
and the dual model once more factorizes as $ D\times T^2$.

In the next section we will extend this duality to the case in which
boundaries are present. As already discussed in \cite{Sevrin:2007yn}, the
main difficulty is the construction of the right boundary terms such that
the boundary conditions of the various fields remain consistent with the
duality transformation. A crucial partial integration identity is here,
\begin{eqnarray}
 &&\int d^2 \sigma\, d^2 \theta \, D' \hat D'\,\big(
i\,u\,\bar \ID\bar \ID 'Y+i\,\bar u\,\ID\ID'Y
 \big)-i\,\int d \tau \,d^2\theta \,\big(
\bar \ID'u \,\bar \ID 'Y-\ID '\bar u\,\ID 'Y
 \big)=\nonumber\\
 &&\qquad \int d^2 \sigma\, d^2 \theta \, D' \hat D'\,\,Y\big(z + \bar z\big).
 \label{pi1}
\end{eqnarray}
where $Y$ is a real and $u$ ($\bar u=u^\dagger$) a complex
unconstrained superfield. We also introduced the chiral field
$z\equiv i\,\bar \ID\bar \ID'u$ and $\bar z\equiv i\, \ID \ID'\bar
u$. Another essential equation is,
\begin{eqnarray}
 &&\int d^2 \sigma\, d^2 \theta \, D' \hat D'\,\big(
i\,u\,\bar \ID_+ \ID_- Y+i\,\bar u\,\ID_+\bar \ID_-Y
 \big)-i\,\int d \tau \,d^2\theta \,\big(
u \,\bar \ID_+ \ID_-Y-\bar u\,\ID_+\bar \ID_-Y
 \big)=\nonumber\\
 &&\qquad \int d^2 \sigma\, d^2 \theta \, D' \hat D'\,\,Y\big(w + \bar w\big)-
\int d \tau \,d^2\theta \,Y\,\big(w-\bar w\big),\label{pi2}
\end{eqnarray}
where $Y$ is a real and $u$ ($\bar u=u^\dagger$) a complex
unconstrained superfield and where we introduced the twisted chiral
field $w\equiv i\,\bar \ID_+\ID_- u$ and $\bar w\equiv i\, \ID_+
\bar \ID_-\bar u$.

Throughout the remainder we will focus on the non-linear $\sigma $-model
on $T^4$ and $S^3\times S^1$ which we believe is sufficient to cover all
subtleties involved. Additional examples can easily be constructed.

\subsection{Dualizing D3-branes on $T^4$}
\subsubsection{Dualizing a chiral field}
We start from the example developed in section \ref{subsubsection:t4}
where we will dualize the chiral field to a twisted chiral field. Without
altering the boundary potential -- as the required generalized K{\"a}hler
transformation yields a total derivative contribution to the boundary
term -- we take for the K{\"a}hler potential $V=-w\bar w+(z+\bar z)^2/2$.
Furthermore we assume that the arbitrary function $f$ in the boundary
potential eq.~(\ref{t45}) exhibits the isometry as well: $f=f(z+\bar z)$.
Our starting point is the first order action,
\begin{eqnarray}
 &&{\cal S}_{(1)}=\int d^2 \sigma\, d^2 \theta \, D' \hat D'\,\left(w \bar w-\frac 1 2\, Y^2+
i\,u\,\bar \ID_+ \ID_- Y+i\,\bar u\,\ID_+\bar \ID_-Y
 \right)\nonumber\\
 &&+i\,\int d \tau \,d^2\theta \,\left(
 \frac{i}{2}\frac{\alpha }{\bar \alpha }\,w^2-\frac{i}{2}\frac{\bar \alpha }{\alpha }\,\bar w^2+f(Y)-
u \,\bar \ID_+ \ID_-Y+\bar u\,\ID_+\bar \ID_-Y
 \right),\label{t51}
\end{eqnarray}
with $Y\in\IR$ and $u\in\IC$ unconstrained superfields. From
eqs.~(\ref{t42})-(\ref{t44}) we obtain the boundary conditions,
\begin{eqnarray}
&& \ID\left( \alpha  w+\bar \alpha\bar w-\beta \,Y\right)=0,\nonumber\\
&& \bar \ID\left( \alpha  w+\bar \alpha\bar w-\bar \beta \,Y\right)=0,\label{t52}
\end{eqnarray}
and,
\begin{eqnarray}
 \ID'Y&=&\ID\left(- \frac{\bar \beta }{\bar \alpha }\, w+\frac{\bar \beta }{\alpha }\,\bar w+i\,f'(Y)\right),\nonumber\\
 \bar \ID'Y&=&\bar \ID\left(+ \frac{\beta }{\bar \alpha }\, w-\frac{ \beta }{\alpha }\,\bar w-i\,f'(Y)\right),
 \label{t53}
\end{eqnarray}
where $f'(Y)=df/dY$. Integrating over $u$ and $\bar u$ in eq.~(\ref{t51})
yields the original model back. Integrating the first order action
eq.~(\ref{t51}) by parts using eq.~(\ref{pi2}) gives,
\begin{eqnarray}
 &&{\cal S}_{(1)}=\int d^2 \sigma\, d^2 \theta \, D' \hat D'\,\left(w \bar w-\frac 1 2\, Y^2+
\big(w'+\bar w'\big)\,Y\right)
 \nonumber\\
 &&+i\,\int d \tau \,d^2\theta \,\left(
 \frac{i}{2}\frac{\alpha }{\bar \alpha }\,w^2-\frac{i}{2}\frac{\bar \alpha }{\alpha }\,\bar w^2+f(Y)+
 i\,\big(w'-\bar w'\big)\,Y\right),\label{t55}
\end{eqnarray}
with $w'$ another twisted chiral field. Integrating over $Y$ gives the
dual model. The bulk equation of motion for $Y$ is,
\begin{eqnarray}
 Y=w'+\bar w'.\label{t56}
\end{eqnarray}
The variation of the boundary term requires more care. Before doing this
we note that we can distinguish two cases: $\beta =\bar \beta $ and
$\beta \neq\bar \beta $. Indeed when $b\equiv \beta $ is real,
eq.~(\ref{t52}) implies a Dirichlet boundary condition,
\begin{eqnarray}
 \alpha w+\bar \alpha \bar w=b\,Y,\label{t54}
\end{eqnarray}
which is no longer true if $\beta \neq \bar \beta $ where $\alpha w+\bar
\alpha \bar w-\beta \,Y$ becomes a complex boundary chiral field. When
$\beta =\bar \beta $ we have -- as can be seen from eq.~(\ref{t42}) -- no
Dirichlet boundary condition in the direction in which we dualize so we
expect a D2-brane in the dual theory. For $\beta \neq \bar \beta $ the
Dirichlet boundary condition eq.~(\ref{t42}) does depend on the direction
in which we dualize resulting in a D4-brane in the dual theory. As the
dual model describes an A-brane on a 4-dimensional
K{\"a}hler manifold, the dual D-brane must be a space-filling coisotropic brane.\\

\vspace{.3cm}

\noindent\underline{i. $b=\beta=\bar\beta$}\\
When $b\neq0$, we get that because of eq.~(\ref{t54}) the variation of
$Y$ is related to that of $w$ and $\bar w$. Taking the boundary
contribution of the variation of $w$ and $\bar w$ into account -- e.g.
using eq.~(\ref{bsfab}) -- we get that the variation of the boundary term
vanishes provided that the Dirichlet boundary condition,
\begin{eqnarray}
 w'-\bar w'=- \frac{b }{\bar \alpha }\, w+\frac{b }{\alpha }\,\bar w+i\,f'(Y),\label{tuy66}
\end{eqnarray}
holds. This is -- using eq.~(\ref{t56}) -- indeed equivalent to
eq.~(\ref{t53}).

The dual action becomes,
\begin{eqnarray}
 &&{\cal S}_{dual}=\int d^2 \sigma\, d^2 \theta \, D' \hat D'\,\left(w \bar w+w'\bar w'\right)
 \nonumber\\
 &&+i\,\int d \tau \,d^2\theta \,\left(
 f(w'+\bar w')-\frac 1 2 \,\big(w'+\bar w'\big)f'(w'+\bar w')\right).\label{t57}
\end{eqnarray}
We get two Dirichlet boundary conditions,
\begin{eqnarray}
&&\alpha w+\bar \alpha \bar w=b\,w'+b\,\bar w',\nonumber\\
&& w'-\bar w'=- \frac{b }{\bar \alpha }\, w+\frac{b
}{\alpha }\,\bar w+i\,f'(w'+\bar w'),
\end{eqnarray}
where the first one follows from eq.~(\ref{t54}) and the second one from
eq.~(\ref{tuy66}). They imply -- because of eq.~(\ref{btcf}) -- two more
Neumann boundary conditions. We end up with a D2-brane (with a flat
$U(1)$ bundle) wrapping around a lagrangian submanifold of $T^4$.

\vspace{.3cm}

\noindent\underline{ii. $\beta\neq\bar\beta$} \\
When performing the variation of the boundary term, one needs to realize
that because of eq.~(\ref{t52}), $\alpha  w+\bar \alpha\bar w-\bar \beta
\,Y$ is a complex chiral field on the boundary. Solving this in terms of
unconstrained superfields and appropriately taking the boundary
contributions of the bulk variations of $w$ and $\bar w$ into account we
get that the variation of the boundary term vanishes provided,
\begin{eqnarray}
&& \ID\left(+w'-\bar w'+\frac{\bar \beta }{\bar \alpha }\, w-\frac{\bar \beta }{\alpha }\,
\bar w-i\,f'(Y)\right)=0,\nonumber\\
&& \bar \ID\left(-w'+\bar w'- \frac{\beta }{\bar \alpha }\, w+\frac{ \beta }{\alpha }\,\bar w+i\,f'(Y)\right)=0.
\end{eqnarray}
This is indeed consistent with eqs.~(\ref{t53}) and (\ref{t56}).

Summarizing, the dual action is given by,
\begin{eqnarray}
 &&{\cal S}_{dual}=\int d^2 \sigma\, d^2 \theta \, D' \hat D'\,\left(w \bar w+w'\bar w'\right)
 \nonumber\\
 &&+i\,\int d \tau \,d^2\theta \,\left(
 \frac i 2 \frac{\alpha }{\bar \alpha }\, w^2-\frac i 2 \frac{\bar \alpha }{\alpha }\,\bar w^2+\frac i 2 \, w'{}^2
 -\frac i 2 \bar w'{}^2+ f(w'+\bar w')\right),
\end{eqnarray}
and the Neumann boundary conditions can be rewritten as,
\begin{eqnarray}
 \hat D w&=&i\,\frac{\beta +\bar \beta }{\beta -\bar \beta }\,Dw+i\,
\frac{\alpha \bar \alpha (1-i\,f'')-\beta \bar \beta }{\alpha (\beta -\bar \beta )}\,Dw' -i\,
\frac{\alpha \bar \alpha (1+i\,f'')+\beta \bar \beta }{\alpha (\beta -\bar \beta )}\,D\bar w',\nonumber\\
\hat D w'&=&i\,
\frac{\alpha \bar \alpha (1+i\,f'')-\beta \bar \beta }{\bar \alpha (\beta -\bar \beta )}\,Dw-
i\,\frac{\beta +\bar \beta }{\beta -\bar \beta }\,Dw'+i\,
\frac{\alpha \bar \alpha (1+i\,f'')+\beta \bar \beta }{\alpha (\beta -\bar \beta )}\,D\bar w,
\end{eqnarray}
together with the complex conjugate of these expressions. We wrote $f''$
for $\partial^ 2 f(w'+\bar w' )/\partial w'{}^2$. Denoting the twisted
chiral fields $w$,$\bar w$, $w'$ and $\bar w'$ collectively by $w^a$, we
can rewrite the Neumann boundary conditions as,
\begin{eqnarray}
 \hat D w^a=K^a{}_bDw^b,
\end{eqnarray}
where consistency with $\hat D^2=-(i/2)\partial /\partial \tau $ requires
that $K$ is a complex structure \cite{Sevrin:2007yn}. Indeed one shows
that the Nijenhuis-tensor of the complex structure vanishes due to the
symmetry $w'+\bar w'$ in the function $f$. Note that $K$ does not
anticommute with the original complex structure. So we get here a
maximally coisotropic D4-brane. Using eq.~(\ref{n2tcsf}) and comparing
the result with eq.~(\ref{nbc9}), we obtain the $U(1)$ fieldstrength,
\begin{eqnarray}
 &&F_{ww'}=\frac{\alpha \bar \alpha (1-i\,f'')+\beta \bar \beta }{\bar \alpha (\beta -\bar \beta )},\quad
 F_{w\bar w}=-\frac{\beta +\bar \beta }{\beta -\bar \beta },\quad
 F_{w\bar w'}=-\frac{\alpha \bar \alpha (1+i\,f'')-\beta \bar \beta }{\bar \alpha (\beta -\bar \beta )},\nonumber\\
 &&F_{w'\bar w}=-\frac{\alpha \bar \alpha (1-i\,f'')-\beta \bar \beta }{\alpha (\beta -\bar \beta )},\quad
 F_{w'\bar w' }=\frac{\beta +\bar \beta }{\beta -\bar \beta },\quad
 F_{\bar w\bar w'}=-\frac{\alpha \bar \alpha (1+i\,f'')+\beta \bar \beta }{\alpha (\beta -\bar \beta )}.\nonumber\\
&&\end{eqnarray} This  generalizes some of the configurations studied in
e.g.~\cite{Kapustin:2001ij}, \cite{Aldi:2005hz}, \cite{Font:2006na} and
\cite{Sevrin:2007yn}.

\subsubsection{Dualizing a twisted chiral field}
With a generalized K{\"a}hler transformation, we can make the isometry
manifest in the bulk potential eq.~(\ref{van1}),
\begin{eqnarray}
 V=z\bar z-\frac 1 2 \big(w+\bar w\big)^2,
\end{eqnarray}
which because of eq.~(\ref{genKahlbd}) modifies the boundary potential
eq.~(\ref{t45}) to,
\begin{eqnarray}
 W=\frac i 2 \,\frac{\alpha +\bar \alpha }{\alpha \bar \alpha }\,\big(\alpha\, w^2-\bar \alpha\, \bar w^2\big)+
 f(z,\bar z). \label{van3}
\end{eqnarray}
The fact that the boundary potential does not manifestly reflect the
isometry -- in fact using the boundary condition eq.~(\ref{t42}) one
shows that it is invariant modulo total derivative terms -- constitutes
the whole subtlety for this particular case. We can rewrite the boundary
condition eq.~(\ref{t42}) as,
\begin{eqnarray}
 \big(\alpha +\bar \alpha \big)\ID(w+\bar w)+\big(\alpha -\bar \alpha \big)\ID'(w+\bar w)=2\beta \,\ID z,\label{van7}
\end{eqnarray}
together with its complex conjugate. Once more we have to distinguish
between two cases: either $\alpha\in \IR $ or $\alpha \neq \bar \alpha $.
In the former case the dual theory describes a D2-brane while for the
latter we will obtain a D4-brane. Finally we can also rewrite the Neumann
boundary conditions in eq.~(\ref{t44}) in an invariant way,
\begin{eqnarray}
 2\alpha \bar \alpha\, \ID'z+\bar \beta \big(\alpha +\bar \alpha \big)\ID'(w+\bar w)=
 -\bar\beta \big(\alpha -\bar \alpha \big)\ID(w+\bar w)+2i\alpha \bar \alpha\,f_{z\bar z}\,\ID z,\label{van4}
\end{eqnarray}
and its complex conjugate.

\vspace{.3cm}

\noindent\underline{i. $a\equiv\alpha=\bar\alpha$}\\
\noindent We first rewrite the boundary potential. Writing $z=\bar
\ID\Lambda $ and $\bar z=\ID\bar \Lambda $ we get,
\begin{eqnarray}
 {\cal S}_{boundary}=i\,\int d \tau \,d^2\theta \,\left\{
-\frac{i\beta }{a}\,\Lambda \,\bar \ID'\big(w+\bar w\big)
+\frac{i\bar \beta }{a}\,\bar \Lambda \, \ID'\big(w+\bar w\big)
+f(z,\bar z) \right\}.
\end{eqnarray}
Using this we write the  first order action,
\begin{eqnarray}
 &&{\cal S}_{(1)}=\int d^2 \sigma\, d^2 \theta \, D' \hat D'\,\left\{-z \bar z+\frac 1 2\, Y^2-
i\,u\,\bar \ID\bar  \ID' Y-i\,\bar u\,\ID \ID'Y
 \right\}
 \nonumber\\
 &&\qquad +i\,\int d \tau \,d^2\theta \,\left\{
-\frac{i\beta }{a}\,\Lambda \,\bar \ID'Y
+\frac{i\bar \beta }{a}\,\bar \Lambda \, \ID'Y
+f(z,\bar z) \right\},\label{van5}
\end{eqnarray}
where $u$ is an unconstrained complex superfield and $Y$ is a real
unconstrained superfield. Varying $u$ yields back the original model.
Varying $z$, $\delta z=\bar\ID\delta  \Lambda $ gives a boundary term
which vanishes provided,
\begin{eqnarray}
 a\,\ID'z+\bar \beta \,\ID'Y=+i\,a\,f_{z\bar z}\,\ID z,
\end{eqnarray}
which is consistent with eq.~(\ref{van4}). Integrating eq.~(\ref{van5})
by parts,
\begin{eqnarray}
 &&{\cal S}_{(1)}=\int d^2 \sigma\, d^2 \theta \, D' \hat D'\,\left\{-z \bar z+\frac 1 2\, Y^2-
Y\big(z'+ \bar z'\big)
 \right\}
 \nonumber\\
 &&\qquad +i\,\int d \tau \,d^2\theta \,\left\{
\bar \ID'Y\left(\frac{i\beta }{a}\,\Lambda +\bar \ID' u\right)-
\ID'Y\left(\frac{i\bar \beta }{a}\,\bar \Lambda +\ID'\bar u\right)
+f(z,\bar z) \right\}.\label{van6}
\end{eqnarray}
Varying $Y$ gives a bulk equation of motion,
\begin{eqnarray}
 Y=z'+\bar z',
\end{eqnarray}
and a boundary contribution which vanishes provided,
\begin{eqnarray}
 i\beta \,\Lambda +a\,\bar \ID' u=i\bar \beta \bar \Lambda +a\,\ID'\bar u=0.
\end{eqnarray}
This immediately implies the Dirichlet boundary conditions,
\begin{eqnarray}
z'=\frac{\beta }{a}\,z,\qquad \bar z'=\frac{\bar \beta }{a}\,\bar z,\label{van9}
\end{eqnarray}
which is consistent with eq.~(\ref{van7}). Combining everything, we get
the dual action,
\begin{eqnarray}
 &&{\cal S}_{dual}=\int d^2 \sigma\, d^2 \theta \, D' \hat D'\,\left(-z \bar z-\frac 1 2\, \big(z'+ \bar z'\big)^2
 \right)
+i\,\int d \tau \,d^2\theta \,
f(z,\bar z) ,\label{van8}
\end{eqnarray}
together with the Dirichlet boundary conditions in eq.~(\ref{van9}) and
the Neumann boundary conditions,
\begin{eqnarray}
 a\,\ID'z+\bar \beta\, \ID'z'=ia\,f_{z\bar z}\,\ID z,\qquad
 a\,\bar \ID'\bar z+ \beta\, \bar \ID'\bar z'=-ia\,f_{z\bar z}\,\bar \ID\bar  z.
\end{eqnarray}
We see that the dual model describes a B type D2-brane wrapping on a
holomorphic cycle determined by eq.~(\ref{van9}) and we have a
non-trivial $U(1)$ bundle with non-vanishing potentials,
\begin{eqnarray}
 A_z=+\frac{i}{2}\,f_z,\qquad  A_{\bar z}=-\frac{i}{2}\,f_{\bar z}\,.
\end{eqnarray}

\vspace{.3cm}

\noindent\underline{ii. $\alpha\neq\bar\alpha$}\\
\noindent Using the Dirichlet boundary condition eq.~(\ref{t42}) we
rewrite the boundary potential eq.~(\ref{van3}) in an invariant way,
\begin{eqnarray}
 W=\frac{i\,(\alpha +\bar \alpha )}{2\,(\alpha -\bar \alpha )}\left(
\frac{1}{\alpha \bar \alpha }\,\big(\beta z+\bar \beta \bar z\big)^2-
(w+\bar w)^2
 \right)+f(z,\bar z).
\end{eqnarray}
With this we obtain the first order action,
\begin{eqnarray}
 &&{\cal S}_{(1)}=\int d^2 \sigma\, d^2 \theta \, D' \hat D'\,\left\{-z \bar z+\frac 1 2\, Y^2-
i\,u\,\bar \ID\bar  \ID' Y-i\,\bar u\,\ID \ID'Y
 \right\}
 \nonumber\\
 &&\qquad +i\,\int d \tau \,d^2\theta \,\bigg\{
\frac{i\,(\alpha +\bar \alpha )}{2\,(\alpha -\bar \alpha )}\left(
\frac{1}{\alpha \bar \alpha }\,\big(\beta z+\bar \beta \bar z\big)^2-
Y^2
 \right)+f(z,\bar z)+\nonumber\\
&& \bar \ID'u\left( \bar \ID'Y+ \frac{2\bar \beta }{\alpha -\bar \alpha }\,\bar \ID\bar z-
\frac{\alpha +\bar \alpha }{\alpha -\bar \alpha }\,\bar \ID Y\right)
- \ID'\bar u\left(  \ID'Y- \frac{2 \beta }{\alpha -\bar \alpha }\, \ID z+
\frac{\alpha +\bar \alpha }{\alpha -\bar \alpha }\, \ID Y\right)
  \bigg\}.\nonumber\\\label{van10}
\end{eqnarray}
Integrating over $u$ and $\bar u$ gives us the original model back
together with a boundary term which vanishes provided,
\begin{eqnarray}
 \bar \ID'Y+ \frac{2\bar \beta }{\alpha -\bar \alpha }\,\bar \ID\bar z-
\frac{\alpha +\bar \alpha }{\alpha -\bar \alpha }\,\bar \ID Y=
\ID'Y- \frac{2 \beta }{\alpha -\bar \alpha }\, \ID z+
\frac{\alpha +\bar \alpha }{\alpha -\bar \alpha }\, \ID Y=0,
\end{eqnarray}
which is consistent with eq.~(\ref{van7}). Integrating by parts, we
rewrite eq.~(\ref{van10}) as,
\begin{eqnarray}
 &&{\cal S}_{(1)}=\int d^2 \sigma\, d^2 \theta \, D' \hat D'\,\left(-z \bar z+\frac 1 2\, Y^2-
Y\big(z' +\bar z'\big)
 \right)
 \nonumber\\
 &&\qquad +i\,\int d \tau \,d^2\theta \,\bigg\{
\frac{i}{2\alpha \bar \alpha }\,\frac{\alpha +\bar \alpha }{\alpha -\bar \alpha }\,
\big(\beta z+\bar \beta \bar z\big)^2-i\,\frac{\alpha +\bar \alpha }{\alpha -\bar \alpha }
\Big(\frac 1 2 Y^2-Y\big(z' +\bar z'\big)\Big)\nonumber\\
&&\qquad-i\,\frac{2\bar \beta }{\alpha -\bar \alpha }\,z'\,\bar z
-i\,\frac{2 \beta }{\alpha -\bar \alpha }\,\bar z'\, z+f(z,\bar z)\bigg\}.\label{van12}
\end{eqnarray}
Integrating over $Y$ gives both in the bulk and in the boundary,
\begin{eqnarray}
 Y=z'+\bar z'.
\end{eqnarray}
It is now straightforward to go to the second order expressions. One
finds for the bulk potential of the dual model,
\begin{eqnarray}
 V=z\bar z+z'\bar z',
\end{eqnarray}
and for its boundary potential,
\begin{eqnarray}
 W&=&i\,\frac{\beta \bar \beta }{\alpha \bar \alpha }\,\frac{\alpha +\bar \alpha }{\alpha -\bar \alpha }\,
 z \bar z+i \,\frac{\alpha +\bar \alpha }{\alpha -\bar \alpha }\,
z '\bar z'
-i\,\frac{2\bar \beta }{\alpha -\bar \alpha }\,z'\,\bar z
-i\,\frac{2 \beta }{\alpha -\bar \alpha }\,\bar z'\, z+f(z,\bar z).
\end{eqnarray}
The boundary conditions are given by,
\begin{eqnarray}
\ID'z&=&-\frac{\beta \bar \beta }{\alpha \bar \alpha }\,\frac{\alpha +\bar \alpha }{\alpha -\bar \alpha }\,\ID z+
i\,f_{z\bar z}\,\ID z+ \frac{2\bar \beta }{\alpha -\bar \alpha }\ID z',
\nonumber\\
 \ID'z'&=& \frac{2\beta }{\alpha -\bar \alpha }\,\ID z -
\frac{\alpha +\bar \alpha }{\alpha -\bar \alpha }\ID z',
\end{eqnarray}
and their complex conjugates. We end up with a B type D4-brane wrapping
around the four torus. Once more we have a non-trivial $U(1)$ bundle with
potentials given by $A_z=i\,W_z/2$, $A_{z' }=i\,W_{z'}/2$ and their
complex conjugates.

\subsection{Dualizing the branes on $S^3\times S^1$}
\subsubsection{Dualizing a twisted chiral field}
\noindent\underline{i. From a D1- to a D2-brane}\\
Our starting point is the D1-brane configuration on $S^3\times S^1$
discussed in section \ref{subsection:S3S1}. For simplicity we choose the
Dirichlet boundary conditions for the chiral field as $z=\bar z=0$. We
start from the first order action,
\begin{eqnarray}
{\cal S}_{(1)}&=&\int d^2 \sigma\, d^2 \theta \, D' \hat D'\,
\bigg\{ \int^{e^Y}\frac{dq}{q}\ln\big(q+z\bar z\big)+i\,u\, \bar \ID
\bar \ID 'Y+ i \,\bar u\, \ID \ID'Y \bigg\}
\nonumber\\
&& + i\int d \tau \, d^2\theta\,\Big\{ Q(Y)-Y\,Q'(Y)- \bar
\ID'u\big(\bar \ID'Y+i\, \bar \ID Q'(Y)\big) +\ID'\bar u\big(
\ID'Y-i\, \ID Q'(Y)\big) \Big\},\nonumber\\\label{evto0}
\end{eqnarray}
where the Lagrange multipliers $u$ and $\bar u=u^\dagger$ are
unconstrained complex superfields. Varying $u$ and $\bar u$ gives
the bulk equations of motion $\bar \ID \bar \ID 'Y=\ID \ID'Y=0$
which are solved by putting $Y=\ln w \bar w$ with $w$ a twisted
chiral superfield. The variation of the Lagrange multipliers yields
a boundary term as well which vanishes if $Y$ satisfies the boundary
conditions,
\begin{eqnarray}
\ID'Y&=&+i\, Q''(Y)\ID Y,\nonumber\\
\bar \ID'Y&=&-i\, Q''(Y)\bar \ID Y.\label{evto1}
\end{eqnarray}
Going to the second order action we precisely recover the D1-brane
discussed in section \ref{subsection:S3S1} and eq.~(\ref{evto1}) is
equivalent to the last boundary condition in eq.~(\ref{Eva4}).

Upon integration by parts we rewrite the first order action
eq.~(\ref{evto0}) as,
\begin{eqnarray}
 {\cal S}_{(1)}&=&\int d^2 \sigma\, d^2 \theta\, D' \hat D'\,
\bigg\{ \int^{e^Y}\frac{dq}{q}\ln\big(q+z\bar z\big)+Y\,\ln z'\bar z'\bigg\}\nonumber\\
&& + i\int d \tau \, d^2\theta\,\Big\{ Q(Y)-Y\,Q'(Y)- Q'(Y)\,\ln
z'\bar z' \Big\},\label{evto2}
\end{eqnarray}
where we introduced the chiral field $ z'$,
\begin{eqnarray}
 \ln z'\equiv i\,\bar  \ID\bar \ID'u,\qquad \ln\bar  z'\equiv i\, \ID\ID'\bar u.
\end{eqnarray}
Varying $Y$ in eq.~(\ref{evto2}) gives an explicit expression for
$Y$,
\begin{eqnarray}
 Y=\ln\big(1- z'' \bar z''\big) -\ln z' \bar z'.\label{evto3}
\end{eqnarray}
Note that the boundary term in the variation vanishes as well by virtue
of eq.~(\ref{evto3}). Going to second order gives the dual model,
 \begin{eqnarray}
 {\cal S}_{dual}&=&\int d^2 \sigma\, d^2 \theta \, D' \hat D'\,
 \left\{\int^{z''\bar z''}\frac{dq}{q}\ln\big(1-q\big)-\frac 1 2 \left(\ln z' \bar z'\right)^2\right\}\nonumber\\
&& + i\int d \tau \,d^2\theta  \,Q(-\ln z'\bar z'),
\end{eqnarray}
Where we redefined $z''\equiv z\,z'$. We have the boundary conditions,
\begin{eqnarray}
&&z''= \bar z''=0, \nonumber\\
&&\ID'z'=+i\,Q'' (-\ln z'\bar z')\,\ID z',\qquad
\bar \ID'\bar z'=-i\,Q'' (-\ln z'\bar z')\,\bar \ID \bar z',
\end{eqnarray}
where the Neumann boundary conditions follow from combining
eqs.~(\ref{evto1}) with (\ref{evto3}).

The dual model has a target geometry given by $D\times T^2$ with a
D2-brane wrapping around the torus.

\vspace{.3cm}
\noindent\underline{ii. From a D3- to a D4-brane}\\
We now turn to the dualization of the D3-brane configuration discussed in
section \ref{subsection:S3S1}. We consider the configuration given by the
two Neumann boundary conditions eq.~(\ref{EV3}) for the chiral superfield
and the Dirichlet boundary condition and Neumann boundary condition
resulting from eq.~(\ref{EV2}) for the twisted chiral superfield. We
start from the first order action,
\begin{eqnarray}
{\cal S}_{(1)}&=&\int d^2 \sigma\, d^2 \theta \, D' \hat D'\,
\bigg\{ \int^{e^Y}\frac{dq}{q}\ln\big(q+z\bar z\big)+i\,u\, \bar \ID
\bar \ID 'Y+ i \,\bar u\, \ID \ID'Y \bigg\}
\nonumber\\
&&+ i\int d \tau \, d^2\theta\,\bigg\{ W(Y)- \bar
\ID'u\Big(\bar \ID'Y+i\, m_1 \, \bar \ID \ln\big( e^Y + z \bar z\big) + i\, m_2\, \bar\ID\, y\Big) \nonumber \\
&&\qquad \qquad +\ID'\bar u\Big( \ID'Y-i\, m_1 \, \ID \ln\big( e^Y
+z
\bar z \big) -i\, m_2\, \ID\, y \Big) \bigg\},
\label{evto4}
\end{eqnarray}
where $W(Y)$ stands for,
\begin{eqnarray}
 W(Y)=-\frac{m_1}{2}\,\ln\big(z\bar z+e^Y\big)-m_2\,y\,\ln z\bar z.
\end{eqnarray}
We introduced the Lagrange multipliers $u$ and $\bar u=u^\dagger$ as
unconstrained complex superfields, just like in the previous section. But
now the gauge field $Y$ also has to satisfy the boundary conditions
following from eq.~(\ref{EV3}),
\begin{eqnarray}
\ID ' \ln z \bar z &=& +i\, m_1\, \ID \ln\left(e^Y + z \bar z \right) -
\frac{m_2}{z \bar z}\, \ID \, e^Y ,\nonumber\\
\bar \ID ' \ln z \bar z &=& -i\, m_1\, \bar \ID \ln\left(e^Y + z
\bar z \right) - \frac{m_2}{z \bar z}\, \bar \ID \, e^Y.
\label{evto5}
\end{eqnarray}
Varying $u$ and $\bar u$ yields the bulk equations of motion $\bar
\ID \bar \ID 'Y=\ID \ID'Y=0$, and a vanishing boundary term if $Y$
satisfies the boundary conditions,
\begin{eqnarray}
\ID'Y&=&+i\, m_1\,  \ID \ln\big(e^Y + z \bar z\big) + i\, m_2\, \ID\, y, \nonumber\\
\bar \ID'Y&=&-i\, m_1\, \bar \ID \ln\big(e^Y + z \bar z\big) -i\,
m_2\, \bar \ID\, y.\label{evto6}
\end{eqnarray}
The bulk equations of motion can be solved by requiring $Y=\ln(w
\bar w)$, where $w$ is a twisted chiral superfield. Implementing
this in the first order action eq.~(\ref{evto4}) we recover the
original model with the D3-brane from section \ref{subsection:S3S1},
for which the boundary conditions eq.~(\ref{evto6}) are equivalent
to the Dirichlet condition eq.~(\ref{EV2}) and the boundary
conditions eq.~(\ref{evto5}) reduce to eq.~(\ref{EV3}).

Using eq.~(\ref{pi1}) to partially integrate the first order action
eq.~(\ref{evto4}), one gets
\begin{eqnarray}
 {\cal S}_{(1)}&=&\int d^2 \sigma\, d^2 \theta\, D' \hat D'\,
\bigg\{ \int^{e^Y}\frac{dq}{q}\ln\big(q+z\bar z\big)+Y\,\ln z'\bar z'\bigg\}\nonumber\\
&&+i\int d \tau \, d^2\theta\,\Big\{ W(Y)- \left(m_1\, \ln\big( e^Y
+ z \bar z \big) + m_2\, y\right)\,\ln z'\bar z'
\Big\},\label{evto7}
\end{eqnarray}
with a new chiral field $z'$,
\begin{eqnarray}
 \ln z'\equiv i\,\bar  \ID\bar \ID'u,\qquad \ln\bar  z'\equiv i\, \ID\ID'\bar u.
\end{eqnarray}
The equations of motion in the dual picture follow from varying $Y$ in
eq.~(\ref{evto7}),
\begin{eqnarray}
 Y=\ln\big(1- z'' \bar z''\big) -\ln z' \bar z',\label{evto8}
\end{eqnarray}
in which we introduced $z'' \equiv z\, z'$. Imposing the solution
eq.~(\ref{evto8}) also eliminates the boundary term arising from the
variation of eq.~(\ref{evto7}) with respect to $Y$.  This solution also
allows us to write down the second order action describing the dual
model,
\begin{eqnarray}
 {\cal S}_{dual}&=&\int d^2 \sigma\, d^2 \theta \, D' \hat D'\,
 \left\{\int^{z''\bar z''}\frac{dq}{q}\ln\big(1-q\big)-\frac 1 2 \left(\ln z' \bar z'\right)^2\right\}\nonumber\\
&& + i\int d \tau \,d^2\theta  \,\left\{ \frac{1}{2}\, m_1\, \left(
\ln z' \bar z'\right)^2 + i\, m_2\, \ln\left(\frac{z''}{\bar
z''}\right)\, \ln z'' \bar z'' - i\,m_2\, \ln\left(\frac{z'}{\bar
z'} \right) \ln z'' \bar z'' \right\}. \nonumber \\ \label{evto9}
\end{eqnarray}
Rewriting the boundary conditions eqs.~(\ref{evto5}), (\ref{evto6})
in terms of the chiral fields $z'$, $z''$ using eq.~(\ref{evto8})
leads to the following four Neumann boundary conditions,
\begin{eqnarray}
&& \ID' \ln \left(1- z'' \bar z''\right) = - m_2\, \ID \ln z' \bar
z', \qquad \bar\ID' \ln \left( 1- z'' \bar z'' \right) = - m_2\,
\bar \ID \ln z' \bar z'\, ,
\nonumber \\
&& \ID' \ln z' \bar z' = +i\, m_1\, \ID \ln z' \bar z' - m_2\, \ID
\ln \left( \frac{z''}{\bar z''} \right), \nonumber\\
&&\bar\ID' \ln z' \bar z' = - i\, m_1\, \bar\ID \ln z' \bar z' +
m_2\, \bar \ID \ln\left( \frac{z''}{\bar z''}\right). \label{evto10}
\end{eqnarray}
One can check that these boundary conditions are consistent with
eq.~(\ref{bsfab}) applied to the dual action in eq.~(\ref{evto9}).

It is clear that the target space geometry of the dual model is
described by $D\times T^2$ with a spacefilling D4-brane.  The $U(1)$
bundle on the D4-bane can be found by using e.g. eq.~(\ref{Eva}) and
eq.~(\ref{Eva1}) and leads to the following fieldstrength,
\begin{eqnarray}
&&F_{z' \bar z'} = - i \frac{m_1}{z' \bar z'}, \quad F_{z'' \bar
z''} = 0, \nonumber \\
&&F_{z' \bar z''} = - \frac{m_2}{z' \bar z''}, \quad F_{z'' \bar z'}
= \frac{m_2}{z'' \bar z'}.
\end{eqnarray}

\subsubsection{Dualizing a chiral field}
Our starting point is the D3-brane configuration described in section
\ref{subsection:S3S1}.  The Neumann boundary conditions for the chiral
superfield are given by eq.~(\ref{EV3}), while the twisted chiral
superfield satisfies the Dirichlet condition eq.~(\ref{EV2}) and the
Neumann condition derived from it. We introduce a real unconstrained
gauge superfield $Y$ satisfying the boundary condition,
\begin{eqnarray}
\ID' Y &=& i\, m_1\, \ID \ln\big(e^Y + w \bar w \big) - m_2\, e^{-Y}\, \ID\, w \bar w, \nonumber \\
\bar\ID' Y &=& - i\, m_1\, \bar\ID \ln\big(e^Y + w \bar w \big) -
m_2 e^{-Y}\, \bar\ID\, w \bar w,\label{bcoiso1}
\end{eqnarray}
and,
\begin{eqnarray}
&&\ID' \ln\left(w \bar w\right) = i\, m_1\, \ID \ln\left(e^Y + w \bar w \right) + m_2\, \ID\, Y , \nonumber \\
&&\bar \ID' \ln\left(w \bar w\right) = -i\, m_1\, \bar\ID
\ln\left(e^Y + w \bar w \right) + m_2\, \bar\ID\, Y. \label{bcoiso2}
\end{eqnarray}
This configuration allows us to distinguish two different cases. The
first case appears when $m_2 = 0$, which will lead to the dual
lagrangian D2-brane. The second situation is characterized by $m_2
\neq 0$, so that we can construct the dual coisotropic D4-brane.

\vspace{0.4cm}

\noindent\underline{i. From a D3-brane to a lagrangian D2-brane}\\
Let us start by taking $m_2 = 0$. In that case the boundary conditions
eqs.~(\ref{bcoiso1}) and (\ref{bcoiso2}) simplify and we can deduce the
following Dirichlet boundary condition from eq.~(\ref{bcoiso2}),
\begin{eqnarray}
-i\ln\left(\frac{w}{\bar w} \right)= m_1\, \ln\left(e^Y + w \bar w
\right). \label{bctc1}
\end{eqnarray}
Hence, we can write the first order action as,
\begin{eqnarray}
{\cal S}_{(1)}&=& - \int d^2 \sigma\, d^2 \theta \, D' \hat D'\,
 \Big\{\int^{e^Y}\frac{dq}{q}\ln\bigg(1+\frac{q}{w \bar w}\bigg) -\frac{1}{2} \big(\ln w \bar w \big)^2
\nonumber \\ && \qquad \qquad \qquad \qquad \qquad \qquad - i u\,
\bar\ID_+ \ID_- Y - i \bar u\, \ID_+
\bar\ID_-Y \Big\}  \nonumber \\
&& +
   i \int d \tau \, d^2\theta\, \Big\{-\frac{m_1}{2} \left(\ln\left(w \bar w + e^Y \right)\right)^2 - u\,
   \bar\ID_+ \ID_- Y + \bar u\, \ID_+ \bar\ID_-
 Y \Big\}, \nonumber \\ \label{ctc0}
\end{eqnarray}
again introducing (unconstrained) complex superfields $u$ and $\bar
u=u^\dagger$. The variation of the action eq.~(\ref{ctc0}) with respect
to $u$ and $\bar u$ yields the equations of motion,
\begin{eqnarray}
\ID_- \bar\ID_+ Y|_{\theta' = \hat \theta ' = 0} = 0 = \ID_+
\bar\ID_- Y|_{\theta' = \hat \theta ' = 0}\, ,
\end{eqnarray}
which is solved by $Y = \ln z \bar z$ with $z$ a chiral superfield.
The second order action gives the original model with a D3-brane,
described by the boundary conditions eqs.~(\ref{EV2}), (\ref{EV3}).

To integrate the action eq.~(\ref{ctc0}) by parts we use the identity
eq.~(\ref{pi2}) and obtain,
\begin{eqnarray}
{\cal S}_{(1)}&=&-\int d^2 \sigma\, d^2 \theta \, D' \hat D'\,
 \left\{\int^{e^Y}\frac{dq}{q}\ln\bigg(1+\frac{q}{w \bar w}\bigg) -\frac{1}{2} \big(\ln w \bar w
 \big)^2 - Y \ln s \bar s\right\}  \nonumber \\
&& + i \int d \tau \, d^2\theta\, \left\{-\frac{m_1}{2}
\left(\ln\left(w \bar w + e^Y \right)\right)^2 + i\, Y\,
\ln\left(\frac{s}{\bar s}\right)
\right\} ,\nonumber \\
\label{ctc1}
\end{eqnarray}
where we introduced the twisted chiral superfield s,
\begin{eqnarray}
\ln s \equiv i \, \bar\ID_+ \ID_- u\, , \qquad \ln \bar s \equiv i\,
\ID_+ \bar\ID_- \bar u\, .
\end{eqnarray}
Varying $Y$ in eq.~(\ref{ctc1}) yields the equation of motion,
\begin{eqnarray}
Y = \ln w'' \bar w'' + \ln\big(1- w' \bar w' \big), \label{ctc2}
\end{eqnarray}
for which we performed the following coordinate transformation,
\begin{eqnarray}
w' = \frac{1}{s} \, ,\qquad w'' = s\, w \,.
\end{eqnarray}

The Dirichlet condition eq.~(\ref{bctc1}) implies that a variation of $Y$
at the boundary is related to a variation of $w$ and $\bar w$. Therefore,
the variation of eq.~(\ref{ctc1}) with respect to $Y$ renders a boundary
term supplemented with a boundary contribution of the variation of $w$
and $\bar w$ -- as can be found in e.g. eq.~(\ref{bsfab}). By virtue of
eq.~(\ref{bctc1}) the boundary variation leads to a Dirichlet boundary
condition for $w'$,
\begin{eqnarray}
-i \ln\left(\frac{w'}{\bar w'} \right) = 0. \label{bctc2}
\end{eqnarray}
Using eqs.~(\ref{bctc1}) and (\ref{ctc2}), we can deduce a (second)
Dirichlet boundary condition for $w''$,
\begin{eqnarray}
-i\, \ln\left(\frac{w''}{\bar w''}\right) &=& m_1\, \ln w'' \bar w'',
\label{bctc3}
\end{eqnarray}
which is indeed consistent with eq.~(\ref{bcoiso1}) using the
equation of motion eq.~(\ref{ctc2}).

The dual model is described by the following (second order) action,
\begin{eqnarray}
{\cal S}_{dual}&=&-\int d^2 \sigma\, d^2 \theta \, D' \hat D'\,
 \left\{ \int^{w' \bar w'} \frac{dt}{t} \ln\big(1-t\big) - \frac{1}{2} \big(\ln w'' \bar w'' \big)^2 \right\}
 \nonumber \\
&& + i \int d \tau \, d^2\theta\, \left\{ -\frac{m_1}{2} \big(\ln
w'' \bar w'' \big)^2\right\}.
 \label{ctc3}
\end{eqnarray}

As a check one verifies that the two Dirichlet boundary conditions
eqs.~(\ref{bctc2}) and (\ref{bctc3}) guarantee that the boundary term in
the variation of the action eq.~(\ref{ctc3}) vanishes.

The dual target space geometry corresponds to $D\times T^2$ with a
D2-brane wrapping along one direction in $D$ and one direction in
$T^2$.  On $T^2$ the brane can only wrap in specific (quantized)
directions, given by the integer $m_1$.

\vspace{0.4cm}

\noindent\underline{ii. From a D3-brane to a coisotropic D4-brane}\\
To arrive at a coisotropic D4-brane it is necessary to assume $m_2 \neq
0$, and that in this case the gauge superfield $Y$ satisfies the complete
boundary conditions eqs.~(\ref{bcoiso1}) and (\ref{bcoiso2}). The first
order action now reads,
\begin{eqnarray}
{\cal S}_{(1)}&=& - \int d^2 \sigma\, d^2 \theta \, D' \hat D'\,
 \Big\{\int^{e^Y}\frac{dq}{q}\ln\bigg(1+\frac{q}{w \bar w}\bigg) -\frac{1}{2} \big(\ln w \bar w \big)^2
 \nonumber \\
 && \qquad \qquad \qquad \qquad \qquad \qquad - i\, u\, \bar\ID_+ \ID_-  Y - i\, \bar u\, \ID_+ \bar\ID_- Y \Big\}
 \nonumber \\
 && + i \int d \tau \, d^2\theta\, \Big\{-\frac{m_1}{2} \left(\ln\left(w \bar w + e^Y \right)\right)^2 +
 Y\,\bigg(i \ln\left(\frac{w}{\bar w}\right)  + m_1\,\ln\big(e^Y +w \bar w\big) \bigg)  \nonumber \\
 &&\qquad\qquad\qquad\qquad- u\, \bar\ID_+ \ID_- Y + \bar u\, \ID_+
 \bar\ID_-Y
 \Big\}.  \label{ctc4}
\end{eqnarray}
Variation of the unconstrained superfields $u$, $\bar u$ allows us
to go back to the original model with a D3-brane, like we mentioned
above.

In order to find the coisotropic D4-brane, we need to integrate the
action~(\ref{ctc4}) by parts using the identity eq.~(\ref{pi2}),
\begin{eqnarray}
{\cal S}_{(1)}&=&-\int d^2 \sigma\, d^2 \theta \, D' \hat D'\,
 \left\{\int^{e^Y}\frac{dq}{q}\ln\bigg(1+\frac{q}{w \bar w}\bigg) -\frac{1}{2} \big(\ln w \bar w
 \big)^2 - Y \ln s \bar s\right\}  \nonumber \\
&& + i \int d \tau \, d^2\theta\, \Big\{-\frac{m_1}{2}
\left(\ln\left(w \bar w + e^Y \right)\right)^2 \nonumber \\
&& \qquad \qquad \qquad \qquad +
 Y\,\bigg(i \ln\left(\frac{w}{\bar w}\right)  + m_1\,\ln\big(e^Y +w \bar w\big) \bigg) +
i\, Y\, \ln\left(\frac{s}{\bar s}\right)
\Big\}. \nonumber \\
\label{ctc5}
\end{eqnarray}
Varying $Y$ in eq.~(\ref{ctc5}) gives the same bulk equations of motion
eq.~(\ref{ctc2}) as above. From the boundary condition
eq.~(\ref{bcoiso2}) we notice that $Y$, $w$ and $\bar w$ are constrained
at the boundary and need to be solved in terms of unconstrained complex
superfields if we want to have the correct boundary variation. One can
show that the variation of the boundary terms, including boundary
contributions of the bulk variation, vanishes provided,
\begin{eqnarray}
\ID \left( \ln\left(\frac{w''}{\bar w''}\right) - i\, m_1\,\ln w'' \bar w'' - m_2\,\ln\left(1-w'
\bar w' \right) \right) = 0, \nonumber \\
\bar\ID\left( \ln\left(\frac{w''}{\bar w''}\right) - i\, m_1\,\ln
w'' \bar w'' + m_2\,\ln\left(1-w' \bar w' \right) \right) = 0.
\label{ctc6}
\end{eqnarray}
which is indeed consistent with eqs.~(\ref{ctc2}), (\ref{bcoiso1})
and (\ref{bcoiso2}). The other two Neumann boundary conditions can
then be derived from eqs.~(\ref{bcoiso2}) and (\ref{ctc2}),
\begin{eqnarray}
\ID \left(\ln\left(\frac{w'}{\bar w'}\right) - m_2\,\ln w''\bar w''
\right) = 0, \nonumber \\
\bar\ID \left(\ln\left(\frac{w'}{\bar w'}\right) + m_2\,\ln w''\bar
w'' \right) = 0. \label{ctc7}
\end{eqnarray}
The dual model is given by the second order action,
\begin{eqnarray}
{\cal S}_{dual}&=&-\int d^2 \sigma\, d^2 \theta \, D' \hat D'\,
 \left\{ \int^{w' \bar w'} \frac{dt}{t} \ln\big(1-t\big) - \frac{1}{2} \big(\ln w'' \bar w'' \big)^2 \right\}
 \nonumber \\
&&+ i \int d \tau \, d^2\theta\, \Big\{ \frac{m_1}{2} \big(\ln w''
\bar w'' \big)^2 + m_1\, \ln\left(1-w'\bar w'\right) \ln\left(
w''\bar w'' \right) \nonumber \\
&& \qquad \qquad + i\,\ln \left(w'' \bar w''\right)
\ln\left(\frac{w''}{\bar w''}\right) + i\, \ln\left(1-w'\bar w'
\right)\ln\left(\frac{w''}{\bar w''}\right) \Big\}.
\end{eqnarray}
One can check the consistency of the dual model by showing that the
boundary term eq.~(\ref{bsfab}) vanishes. However, the boundary
conditions eqs.~(\ref{ctc6}) and (\ref{ctc7}) imply that $w'$ and
$w''$ are constrained at the boundary and that the boundary
conditions can be solved by introducing chiral boundary superfields.
Taking the variation to these boundary fields yields a boundary term
which vanishes by virtue of eq.~(\ref{ctc6}).

 The Neumann boundary conditions eqs.~(\ref{ctc6}) and
(\ref{ctc7}) may be written as,
\begin{eqnarray}
\hat{D} w' &=& i\, \frac{m_2^2 w'\bar w' - (1-i\,m_1\,)(1-w'\bar
w')}{2m_2\, w'' \bar w'}Dw'' +\nonumber\\
&&i \frac{m_2^2 w'\bar w' +
(1+i\,m_1\,)(1-w'\bar w')}{2m_2\, \bar w'' \bar w'}D \bar w''\,, \nonumber \\
\hat{D} w'' &=& -i\, w'' \frac{m_2^2 w'\bar w' - (1+i\,m_1\,)(1-w'\bar
w')}{2m_2\, w' (1-w' \bar w')}Dw' -\nonumber\\
&& i\, w'' \frac{m_2^2 w'\bar w' +
(1+i\,m_1\,)(1-w'\bar w')}{2m_2\, \bar w' (1- w' \bar w')}D\bar w', \label{bcoiso3}
\end{eqnarray}
accompanied by the complex conjugate of these conditions. One can show
that the Nijenhuis-tensor of the complex structure indeed vanishes.
Hence, the dual model is a coisotropic D4-brane on $D\times T^2$ with the
$U(1)$ fieldstrength given by,
\begin{eqnarray}
&&F_{w'w''} = +\frac{m_2^2 w' \bar w' + (1-i\, m_1)(1-w'\bar
w')}{2m_2\, w' w'' (1-w' \bar w')}, \quad F_{w' \bar w''} =
+\frac{m_2^2 w' \bar w'-(1+i\, m_1\,)(1-w' \bar w')}{2 m_2\, w' \bar
w''(1-w' \bar w')}, \nonumber \\
&&F_{w'' \bar w'} =- \frac{m_2^2 w' \bar w' - (1-i\, m_1)(1-w' \bar
w')}{2m_2 \bar w' w'' (1- w' \bar w')}, \quad F_{\bar w' \bar w''}=
+\frac{m_2^2 w' \bar w'+(1+i\, m_1)(1-w' \bar w')}{2m_2 \bar w' \bar
w'' (1-w' \bar w')} \,.\nonumber \\ \label{ctc8}
\end{eqnarray}
This is an interesting example of a maximally coisotropic D-brane as the
target manifold $D\times T^ 2$ is not hyper-K{\"a}hler\footnote{This can
easily be seen from the fact that the K{\"a}hler potential does not satisfy
the Monge-Amp{\`e}re equation, $V_{w'\bar w'}V_{w''\bar w''}- V_{w'\bar
w''}V_{w''\bar w'}=h(w',w'')\bar  h(\bar w',\bar w'' )$
 with $h$ some non-vanishing
holomorphic function.} in contrast with previously studied examples of
coisotropic branes \cite{Kapustin:2001ij}, \cite{Aldi:2005hz},
\cite{Font:2006na} and \cite{Sevrin:2007yn}.

\subsection{Dualizing a chiral/twisted chiral pair to a
semi-chiral multiplet}

While we will discuss D-branes in a semi-chiral geometry in detail in
\cite{wip} we can already gain some insights by using the duality
transformation proposed in \cite{Grisaru:1997ep} which -- if an
appropriate isometry is present -- allows one to dualize a pair
consisting of a chiral and a twisted chiral superfield into a semi-chiral
superfield and vice-versa. In \cite{Lindstrom:2007vc},
\cite{Gates:2007ve}, \cite{Lindstrom:2007sq} and \cite{Merrell:2007sr},
the underlying gauge theory structure has been developed and T-duality
transformations were discussed. We first briefly review the case without
boundaries closely following the treatment in \cite{Lindstrom:2007sq}.
Consider a system described by a single chiral superfield $z$ and a
single twisted chiral superfield $w$. The potential has the form,
\begin{eqnarray}
 V=V\big(z+\bar z,w+\bar w,i(z-\bar z-w+\bar w)\big).
\end{eqnarray}
We introduce three unconstrained real superfields $Y$, $\tilde Y$ and
$\hat Y$ and construct the complex combinations,
\begin{eqnarray}
Y_L\equiv\hat Y+i\,\big(Y-\tilde Y\big),\qquad
Y_R\equiv\hat Y+i\,\big(Y+\tilde Y\big).
\end{eqnarray}
Note that $Y_L$ and $Y_R$ are not independent as $Y_L+\bar Y_L=Y_R+\bar
Y_R$. With this we write down the first order action,
\begin{eqnarray}
 {\cal S}_{(1)}&=& 4\, \int d^2 \sigma\, d^2 \theta\,d^2\hat\theta\,
 \Big\{V(Y,\tilde Y,\hat Y)+\upsilon ^+\,\bar \ID_+Y_L
 +\bar \upsilon ^+ \,\ID_+\bar Y_L+\nonumber\\
&&\qquad \upsilon ^-\,\bar \ID_-Y_R
 +\bar \upsilon ^- \,\ID_-\bar Y_R.
 \Big\}
\end{eqnarray}
Integrating over the unconstrained complex fermionic Lagrange multipliers
$\upsilon ^{\pm}$ and $\bar \upsilon ^{\pm}$ puts the semi-chiral gauge
invariant fieldstrengths to zero: $\IF_+=\bar \IF_+=\IF_-=\bar \IF_-=0$
where,
\begin{eqnarray}
 \IF_+=i\,\bar \ID_+Y_L,\qquad
 \bar \IF_+=i\, \ID_+Y_L,\qquad
 \IF_-=i\,\bar \ID_-Y_R,\qquad
 \bar \IF_-=i\, \ID_-Y_R.
\end{eqnarray}
This is solved by putting $Y_L=2i\,(z-w)$ and $Y_R=2i\,(z+\bar w)$ which
brings us back to the original model. If on the other hand we integrate
over $Y$, $\tilde Y$ and $\hat Y$, we obtain the dual model which is now
a function of the semi-chiral fields $r\equiv \bar \ID_+\upsilon ^+$,
$\bar r\equiv \ID_+\bar \upsilon ^+$, $s\equiv \bar \ID_-\upsilon ^-$ and
$\bar s\equiv \ID_-\bar \upsilon ^-$. They satisfy $\bar \ID_+r=\ID_+\bar
r=\bar \ID_-s=\ID_-\bar s=0$ \cite{Buscher:1987uw}.

We now consider boundaries as well. For concreteness we will start from
the D3-brane on $T^4$, discussed in section 5.1.2, as an explicit
example. For simplicity we choose $\alpha =i\,a\neq 0$ and $\beta =i\,b$,
$a,b\in\IR$, which results in a Dirichlet boundary condition of the form,
\begin{eqnarray}
 -i\,\big(w-\bar w\big)=-i\,\frac b a \big(z-\bar z\big).\label{dir42}
\end{eqnarray}
Using a general K{\"a}hler transformation we write the potential as,
\begin{eqnarray}
&& V\big(z+\bar z,w+\bar w,i(z-\bar z-w+\bar w)\big)= \frac{g+1}{2}\,\big(z+\bar z\big)^2+
 \frac{g-1}{2}\,\big(w+\bar w\big)^2+\nonumber\\
&& \qquad\qquad\frac g 2\, \big(z-\bar z-w+\bar w\big)^2,
\end{eqnarray}
where $g\in\IR$ and $g\not\in \{0,\pm 1\}$. This in its turn implies a
boundary potential,
\begin{eqnarray}
 W\big(z+\bar z,w+\bar w,i(z-\bar z-w+\bar w)\big)=i\,g\,\big(w+\bar w\big)\big(z-\bar z-w+\bar w\big),
\end{eqnarray}
where once more we simplified the expressions by taking $f(z,\bar z)=0$
in eq.~(\ref{t45}). With this we write the first order action,
\begin{eqnarray}
&&{\cal S}_{(1)}=-\int d^2 \sigma\, d^2 \theta \, D' \hat D'\,\Big(
V(Y,\tilde Y,\hat{Y})+\upsilon ^+\,\bar \ID_+Y_L
 +\bar \upsilon ^+ \,\ID_+\bar Y_L+\upsilon ^-\,\bar \ID_-Y_R
 +\bar \upsilon ^- \,\ID_-\bar Y\Big)\nonumber\\
 &&+i\,\int d \tau \,d^2\theta \,\Big(W(Y,\tilde Y,\hat{Y})
 -i\,\upsilon ^+\,\bar \ID_+Y_L
 +i\,\bar \upsilon ^+ \,\ID_+\bar Y_L+i\,\upsilon ^-\,\bar \ID_-Y_R
 -i\,\bar \upsilon ^- \,\ID_-\bar Y_R\Big).
\end{eqnarray}
Integrating over the Lagrange multipliers $\upsilon^{\pm} $ and $\bar
\upsilon^{\pm} $ brings us back to the original model. Integrating by parts,
we rewrite the first order action as,
\begin{eqnarray}
&&{\cal S}_{(1)}=-\int d^2 \sigma\, d^2 \theta \, D' \hat D'\,\Big(
V(Y,\tilde Y,\hat{Y})+
i\,Y\big(r-\bar r+s-\bar s\big)-i\,\tilde Y\big(r-\bar r-s+\bar s\big)
\nonumber\\
&&\qquad \qquad+\hat{Y}\big(r+\bar r+s+\bar s\big)
\Big)
+i\,\int d \tau \,d^2\theta \,\Big(W(Y,\tilde Y,\hat{Y})+
Y\big(r+\bar r-s-\bar s\big)\nonumber\\
&&\qquad \qquad-\tilde Y\big(r+\bar r+s+\bar s\big)
-i\,\hat{Y}\big(r-\bar r-s+\bar s\big)
\Big).\label{dir43}
\end{eqnarray}
The bulk equations of motion for $Y$, $\tilde Y$ and $\hat{Y}$ give,
\begin{eqnarray}
 Y&=&-\frac{i}{g+1}\,\big(r-\bar r+s-\bar s\big),\nonumber\\
 \tilde Y&=&\frac{i}{g-1}\,\big(r-\bar r-s+\bar s\big),\nonumber\\
 \hat{Y}&=&\frac{1}{g}\,\big(r+\bar r+s+\bar s\big),\label{dir44}
\end{eqnarray}
from which we get the dual potential,
\begin{eqnarray}
 V_{dual}(r,\bar r,s,\bar s)&=&\frac{1}{2(g+1)}\,\big(r-\bar r+s-\bar s\big)^2+
 \frac{1}{2(g-1)}\,\big(r-\bar r-s+\bar s\big)^2\nonumber\\
 &&+\frac{1}{2g}\,\big(r+\bar r+s+\bar s\big)^2.\label{dir45}
\end{eqnarray}

The treatment of the boundary terms requires more care. Once more we have
to distinguish two cases: $a=b$ and $a\neq b$. The former will yield a
D2-brane while the latter gives a D4-brane.

\vspace{.3cm}

\noindent\underline{i. $a=b$}\\

\vspace{.1cm}

{From} eq.~(\ref{dir42}) we find that the gauge fields satisfy a
Dirichlet boundary condition,
\begin{eqnarray}
 \hat{Y}=0.\label{dir46}
\end{eqnarray}
Implementing this in the boundary term of eq.~(\ref{dir43}) we find that
integrating over $\tilde Y$ and $Y$ yields two Dirichlet boundary
conditions in the dual model,
\begin{eqnarray}
 r+\bar r=s+\bar s=0,\label{dir47}
\end{eqnarray}
which is consistent with eqs.~(\ref{dir46}) and (\ref{dir44}). As will be
shown in \cite{wip}, a Dirichlet boundary condition on a semi-chiral
superfield is always paired with a Neumann boundary condition, analogous
to what happens for a twisted chiral superfield. So we end up with
D2-brane whose position is determined by eq.~(\ref{dir47}). The dual
generalized K{\"a}hler potential is given by eq.~(\ref{dir45}) and the dual
boundary potential vanishes, $W_{dual}(r,\bar r,s,\bar s)=0$.

\vspace{.3cm}

\noindent\underline{ii. $a\neq b$}\\

\vspace{.1cm}

In order that our expressions are not unnecessarily cluttered we choose
$a=1$ and $b=0$ (other choices yield similar results as long as $a\neq
b$). We find now that eq.~(\ref{dir42}) implies the boundary conditions,
\begin{eqnarray}
 \bar \ID \big(\hat{Y}-i\,Y\big)= \ID \big(\hat{Y}+i\,Y\big)=0.\label{dir48}
\end{eqnarray}
This means that $\hat{Y}-i\,Y$ is a boundary chiral field. Integrating
over $\tilde Y$ in the boundary term of eq.~(\ref{dir43}) gives an
expression equivalent to the last of eq.~(\ref{dir44}). When integrating
over $Y$ and $\tilde Y$ in the boundary term, we need to take the fact
that they are constrained -- as expressed by eq.~(\ref{dir48}) --
properly into account. We find that the variation vanishes provided the
following two Neumann boundary conditions hold,
\begin{eqnarray}
 &&\bar \ID\big(g\,r+(g-2)\bar r-g\,s-(g-2)\bar s\big)=0,\nonumber\\
 &&\ID\big((g-2)r+g\,\bar r-(g-2)s-g\,\bar s\big)=0.\label{dir49}
\end{eqnarray}
Using eq.~(\ref{dir44}) we can write eq.~(\ref{dir48}) in the second
order formalism which yields two more Neumann boundary conditions,
\begin{eqnarray}
 &&\bar \ID\big(r+(1+2g)\bar r+s+(1+2g)\bar s\big)=0,\nonumber\\
 &&\ID\big((1+2g)r+\bar r+(1+2g)s+\bar s\big)=0.\label{dir50}
\end{eqnarray}
So now we end up with a D4-brane. Note that the boundary conditions
eqs.~(\ref{dir49}) and (\ref{dir50}) imply the existence of an additional
complex structure similar to maximally coisotropic branes on $T^4$. The
generalized K{\"a}hler potential is given by eq.~(\ref{dir45}) while the
boundary potential is now given by,
\begin{eqnarray}
&& W_{dual}(r,\bar r,s,\bar s)= -\frac{i}{g(g+1)}\,
 \Big\{
(r+s)\big((1+2g)(r-s)-(\bar r-\bar s)\big)\nonumber\\
&&\qquad\qquad-
(\bar r+\bar s)\big((1+2g)(\bar r-\bar s)-(r-s)\big)
 \Big\}.
\end{eqnarray}

In fact this particular example already perfectly illustrates the two
possible types of boundary conditions one can have when dealing with a
semi-chiral multiplet: either one has 2 Dirichlet and 2 Neumann
conditions or one has 4 Neumann conditions \cite{wip}.

\section{Conclusions and discussion}
We investigated the allowed boundary conditions for a non-linear $\sigma
$-model in $N=2$ boundary superspace parameterized by chiral and twisted
chiral superfields. This corresponds to classifying D-branes in a
bihermitian target manifold geometry for which the two complex structures
commute. There is no need to distinguish between A- and B-type superspace
boundaries as changing the superspace boundary from B-type (which we used
throughout the paper) to A-type amounts to exchanging the chiral
superfields for twisted chiral superfields and vice-versa. Having $n$,
$n\in\IN$ chiral superfields and $2m_1+m_2$, $m_1\in\IN$,
$m_2\in\{0,1\}$, twisted chiral superfields we found that Dp-brane
configurations are possible where $p=2(a+b+m_1)+m_2$ with
$a\in\{0,1,2,\cdots , n\}$ and $b\in\{0,1,2,\cdots , m_1\}$. Whenever
$b\neq 0$ one needs an additional complex structure on (a subspace of)
the target manifold.

In fact after the initial exploration of semi-chiral fields in the
presence of boundaries in section 6.4 we can already anticipate on the
results of \cite{wip} and illustrate the emerging general picture. In
table \ref{table:a} we summarize the various $N=(2,2)$ superfields and
list their components in $N=2$ boundary superspace. Chiral fields give
rise to constrained boundary superfields while twisted chiral and
semi-chiral fields give unconstrained boundary superfields. Looking at
the unconstrained boundary superfields one realizes immediately that
imposing a Dirichlet boundary condition on them implies a Neumann
boundary condition as well. A second type of boundary conditions for the
unconstrained boundary superfields is obtained by requiring that a
certain combination of them becomes chiral on the boundary. For this one
needs an additional complex structure on a subspace of the target
manifold. All these cases were illustrated in the examples developed in
sections 5 and 6.
\begin{table}
  \centering
  \caption{The three types of $N=(2,2)$ superfields together with their reduction to $N=2$
  boundary superspace.}\label{table:a}
  \begin{tabular}{|c|c|c|c|}
    \hline\hline
    $N=(2,2)$ fields & $N=(2,2)$ constraints & $N=2$ fields & boundary type  \\
    \hline\hline
    chiral: $z$, $\bar z$ & $\bar \ID_\pm\bar  z=\ID_\pm z=0$ & $z$, $\ID'z$,
    $\bar z$, $\bar \ID'\bar z$ & constrained \\
    \hline
    twisted chiral: $w$, $\bar w$ & $\bar \ID_+w=\ID_-w=0$,  &
    $w$, $\bar w$ & unconstrained \\ & $ \ID_+\bar w=\bar \ID_-\bar w=0$& &
    \\\hline
    semi-chiral: $r$, $\bar r$, $s$, $\bar s$ & $ \bar \ID_+r= \ID_+\bar r=0$,
      & $r$, $\bar r$, $s$, $\bar s$ & unconstrained, the  \\
    & $\bar \ID_-s= \ID_-\bar s=0$&$\ID' r$, $\bar \ID'\bar r$, $\ID' s$, $\bar \ID'\bar s$
    &last 4 are auxiliary \\\hline\hline
  \end{tabular}
\end{table}

In order to make direct contact with string compactifications we have to
address the study of D-branes in the six dimensional case. With what we
have learned from the previous we find that we can distinguish six
different cases according to their superfield content.
\begin{enumerate}
  \item {\em 3 chiral superfields}\\
  These are B-branes on a K{\"a}hler manifold. We can have D0-, D2-,
  D4- or D6-branes wrapping on holomorphic cycles.
  \item {\em 2 chiral superfields and one twisted chiral superfield}\\
  We can have D1-, D3- or D5-branes on a bihermitian manifold with
  commuting complex structures.
  \item {\em 1 chiral superfield and two twisted chiral superfields}\\
  The manifold is bihermitian with commuting complex structures. It
allows for D2- or D4-branes with the standard boundary conditions for
  the twisted chiral superfields. However, if the manifold allows for
  generalized coisotropic boundary conditions on the twisted chiral
  superfields one gets in addition a new type of D4-branes and
  D6-branes.
  \item {\em 3 twisted chiral superfields}\\
  Here we are dealing with A-branes on K{\"a}hler manifolds. Either we
  have a lagrangian D3-brane or a coisotropic D5-brane.
  \item {\em 1 chiral superfield and a semi-chiral multiplet}\\
  The manifold is bihermitian and the kernel of the commutator of the
  two complex structures is 2-dimensional. If one imposes Dirichlet
  boundary conditions in the semi-chiral directions one can have D2-
  or D4-branes. Having full Neumann boundary conditions in the
  semi-chiral directions gives either D4- or D6-branes.
  \item {\em 1 twisted chiral superfield and a semi-chiral multiplet}\\
  The manifold is bihermitian and the kernel of the commutator of the
  two complex structures is 2-dimensional. If one imposes Dirichlet
  boundary conditions in the semi-chiral directions one can have a
  D3-brane. Having full Neumann boundary conditions in the
  semi-chiral directions gives a D5-brane.
\end{enumerate}
One sees that even in very simple geometries such as tori -- which can be
described in terms of any of the field combinations listed above -- there
is a wealth of D-brane configurations compatible with the $N=2$
supersymmetry. This might have interesting consequences for model
building using intersecting brane configurations (see {\em e.g.}
\cite{Font:2006na} where the use of coisotropic branes in such settings
was initialized).

Note from the discussion above that D0- and D1-branes preserving the
$N=2$ supersymmetry are relatively ``rare''. Indeed D0-branes can only
occur on K{\"a}hler manifolds solely described in terms of chiral fields. On
the other hand we find that $D1$-branes require a target manifold
geometry described in terms of a single twisted chiral and an arbitrary
number of chiral superfields.

The present analysis clearly motivates a thorough study of semi-chiral
superfields in the presence of boundaries as well \cite{wip}. One
potentially interesting approach could be to ``linearize'' the model
along the lines of \cite{Lindstrom:2007qf}. Indeed there it was shown
that any model described in terms of $m$ semi-chiral multiplets is
equivalent to a gauged $\sigma $-model in terms of $2m$ chiral and $2m$
twisted chiral superfields. While the ungauged model has an indefinite
metric, we do not see any obvious obstruction to apply the results
obtained in this paper to this particular instance.

It is clear that it would be desirable to have a better (global)
geometric characterization of these models, {\em e.g.} by combining the
present results with those in \cite{Lindstrom:2002jb} and
\cite{Albertsson:2002qc}. Presumably a formulation in terms of
generalized complex geometry (see {\em e.g.} \cite{Gualtieri:2007ng})
will clarify many issues. Indeed, it has been shown \cite{Zabzine:2004dp}
that the correct generalization of the notion of A and B branes in this
context corresponds to that of a generalized complex submanifold of a
generalized K{\"a}hler manifold. This is presently under investigation.

The study of the duality transformations between chiral and twisted
chiral superfields led to a surprisingly simple method to construct new
examples of coisotropic D-branes. In particular we explicitly constructed
the first example of a coisotropic D-brane on a manifold which is not
hyper-K{\"a}hler. The method developed in the examples can easily be extended
to a general construction. Take e.g. a model with generalized K{\"a}hler
potential given by $V(z+\bar z,w+\bar w)$ and a prepotential $Q(z+\bar
z,w+\bar w)$. We consider a D3-brane with Dirichlet boundary condition,
\begin{eqnarray}
 -i\big(w-\bar w\big)=-\frac{Q' }{V''}-ia\,\big(z-\bar z\big),
\end{eqnarray}
where $a\in\IR$ and where a prime denotes a derivative with respect to
$w$. The boundary potential $W$ is then given by,
\begin{eqnarray}
 W=Q-\frac{Q'V'}{V''}.
\end{eqnarray}
When dualizing the chiral to a twisted chiral field we distinguish to
cases:
\begin{itemize}
  \item $a=0$ resulting in a dual model where a D2-brane wraps a
  lagrangian submanifold.
  \item $a\neq 0$ resulting in a dual model where we have a space filling
  coisotropic D4-brane.
\end{itemize}

Another immediate application of the present results would be an analysis
of the $\beta $-functions for the models discussed. As shown in
\cite{Nevens:2006ht}, such a calculation is greatly facilitated by doing
it in $N=2$ superspace which automatically gives the stability conditions
that are satisfied by supersymmetric configurations. A particularly
simple and straightforward exercise would be the calculation of the
1-loop $\beta $-function for the maximally coisotropic D4-brane on $T^4$
constructed in section 6.2.1 and this would make a direct connection with
the results developed in \cite{Kapustin:2003se}.

Finally, the present analysis could perhaps allow to simplify some of the
results in \cite{Herbst:2008jq} by reformulating the gauged linear
$\sigma $-models in $N=2$ boundary superspace.

\acknowledgments

\bigskip

We thank Matthias Gaberdiel, Ulf Lindstr{\"o}m, Martin Ro\v cek and
Philippe Spindel for useful discussions and suggestions. All authors are
supported in part by the European Commission FP6 RTN programme
MRTN-CT-2004-005104. AS and WS are supported in part by the Belgian
Federal Science Policy Office through the Interuniversity Attraction Pole
P6/11,  and in part by the ``FWO-Vlaanderen'' through project G.0428.06.
AW is supported in part by grant 070034022 from the Icelandic Research
Fund.

\appendix

\section{Conventions, notations and identities}\label{app conv}
We denote the worldsheet coordinates by $ \tau \in\IR$ and $ \sigma \in \IR$, $ \sigma \geq 0$.
Sometimes we use worldsheet light-cone coordinates,
\begin{eqnarray}
\sigma ^\pp= \tau + \sigma ,\qquad \sigma ^== \tau - \sigma .\label{App1}
\end{eqnarray}
The $N=(1,1)$ (real) fermionic coordinates are denoted by $ \theta ^+$ and $ \theta ^-$ and the
corresponding derivatives satisfy,
\begin{eqnarray}
D_+^2= - \frac{i}{2}\, \partial _\pp \,,\qquad D_-^2=- \frac{i}{2}\, \partial _= \,,
\qquad \{D_+,D_-\}=0.\label{App2}
\end{eqnarray}
The $N=(1,1)$ integration measure is explicitely given by,
\begin{eqnarray}
\int d^ 2 \sigma \,d^2 \theta =\int d^2 \sigma \,D_+D_-.
\end{eqnarray}
Passing from $N=(1,1)$ to $ N=(2,2)$ superspace requires
the introduction of two more real fermionic coordinates $ \hat \theta ^+$ and $ \hat \theta ^-$
where the corresponding fermionic derivatives satisfy,
\begin{eqnarray}
\hat D_+^2= - \frac{i}{2} \,\partial _\pp \,,\qquad \hat D_-^2=- \frac{i}{2} \,\partial _= \,,
\end{eqnarray}
and again all other -- except for (\ref{App2}) -- (anti-)commutators do vanish.
The $N=(2,2)$ integration measure is,
\begin{eqnarray}
\int d^2 \sigma \,d^2 \theta \, d^2 \hat \theta =
\int d^2 \sigma \,D_+D_-\, \hat D_+ \hat D_-.
\end{eqnarray}
Quite often a complex basis is used,
\begin{eqnarray}
\ID_\pm\equiv \hat D_\pm+i\, D_\pm,\qquad
\bar \ID_\pm\equiv\hat D_\pm-i\,D_\pm,
\end{eqnarray}
which satisfy,
\begin{eqnarray}
\{\ID_+,\bar \ID_+\}= -2i\, \partial _\pp\,,\qquad
\{\ID_-,\bar \ID_-\}= -2i\, \partial _=,
\end{eqnarray}
and all other anti-commutators do vanish.

When dealing with boundaries in $N=(2,2)$ superspace, we introduce
various derivatives as linear combinations of the previous ones. We
summarize their definitions together with the non-vanishing
anti-commutation relations. We have,
\begin{eqnarray}
&&D\equiv D_++D_-,\qquad \hat D\equiv \hat D_++ \hat D_-, \nonumber\\
&& D'\equiv D_+-D_-,\qquad \hat D'\equiv \hat D_+- \hat D_-,
\end{eqnarray}
with,
\begin{eqnarray}
&&D^2=\hat D^2=D'{}^2=\hat D'{}^2=- \frac{i}{2} \partial _ \tau , \nonumber\\
&&\{D,D'\}=\{\hat D, \hat D'\}=-i \partial _ \sigma .
\end{eqnarray}
In addition we also use,
\begin{eqnarray}
&&\ID\equiv \ID_++\ID_-=\hat D+i\,D,\qquad \ID'\equiv \ID_+-\ID_-=\hat D'+i\,D', \nonumber\\
&& \bar \ID\equiv \bar \ID_++ \bar \ID_-=\hat D-i\,D,\qquad \bar \ID'\equiv \bar \ID_+- \bar
\ID_-=\hat D'-i\,D'.
\end{eqnarray}
They satisfy,
\begin{eqnarray}
&&\{\ID, \bar \ID\}=\{\ID', \bar \ID'\}= -2i\, \partial _ \tau ,\, \nonumber\\
&&\{\ID, \bar \ID'\}=\{\ID', \bar \ID\}=-2i\, \partial _ \sigma  \,.
\end{eqnarray}

\end{document}